\documentclass[journal]{IEEEtran}
\IEEEoverridecommandlockouts

\newtheorem{Remark}{Remark}
\usepackage{algorithm}
\usepackage{algorithmic}
\usepackage{graphicx}
\usepackage{textcomp}
\usepackage{stfloats}
\usepackage{subfigure}
\usepackage{amssymb,amsmath,bm}
\usepackage{cite}
\usepackage{times}
\usepackage{latexsym}
\usepackage{array}
\usepackage{fancyhdr}
\usepackage{psfrag}
\usepackage{multirow}
\usepackage{xcolor}
\usepackage{color}
\usepackage{makecell}
\usepackage{threeparttable}
\usepackage{hyperref} 
\usepackage[OT1]{fontenc}

\newcommand{\blue}[1]{\color{black}#1}

\begin{document} 
\title{Integrated Massive Communication and Target Localization in 6G Cell-Free Networks\\}

\author{\IEEEauthorblockN{Junyuan~Gao,
Weifeng~Zhu,
Shuowen~Zhang,
Yongpeng~Wu, 
Jiannong~Cao, 
Giuseppe~Caire, 
and Liang~Liu
}
\thanks{J. Gao, W. Zhu, S. Zhang, J. Cao, and L. Liu are with the Department of Electrical and Electronic Engineering, The Hong Kong Polytechnic University, Hong Kong SAR (e-mail: \{junyuan.gao, eee-wf.zhu, shuowen.zhang, jiannong.cao, liang-eie.liu\}@polyu.edu.hk).} 
\thanks{Y. Wu is with the Department of Electronic Engineering, Shanghai Jiao Tong University, Minhang 200240, China (e-mail: yongpeng.wu@sjtu.edu.cn).}
\thanks{G. Caire is with the Communications and Information Theory Group, Technische Universit{\"a}t Berlin, Berlin 10587, Germany (e-mail: caire@tu-berlin.de).}
}

\maketitle

\begin{abstract}  

This paper presents an initial investigation into the combination of integrated sensing and communication (ISAC) and massive communication, both of which are largely regarded as key scenarios in sixth-generation (6G) wireless networks. Specifically, we consider a cell-free network comprising a large number of users, multiple targets, and distributed base stations (BSs). In each time slot, a random subset of users becomes active, transmitting pilot signals that can be scattered by the targets before reaching the BSs. Unlike conventional massive random access schemes, where the primary objectives are device activity detection and channel estimation, our framework also enables target localization by leveraging the multipath propagation effects introduced by the targets. However, due to the intricate dependency between user channels and target locations, characterizing the posterior distribution required for minimum mean-square error (MMSE) estimation presents significant computational challenges. To handle this problem, we propose a hybrid message passing-based framework that incorporates multiple approximations to mitigate computational complexity. Numerical results demonstrate that the proposed approach achieves high-accuracy device activity detection, channel estimation, and target localization simultaneously, validating the feasibility of embedding localization functionality into massive communication systems for future 6G networks.

\end{abstract}

\section{Introduction}\label{sec:intro}

\subsection{Motivation}

Integrated sensing and communication (ISAC) has recently received significant attention from both academia~\cite{ref:FLiu-ISAC2} and industry~\cite{ref:indust-ISAC2}. 
By performing sensing and communication functionalities via the same frequency band and hardware, the ISAC technology is anticipated to play a crucial role in many emerging applications such as smart transportation, factory automation, etc. 
As a matter of fact, IMT-2030 Framework has identified ISAC as one of the primary usage scenarios in sixth-generation (6G) networks~\cite{ref:indust-IMT}. 
{\blue In practical ISAC scenarios, the wireless propagation environment typically comprises multiple targets, such as traffic signs, billboards, and vehicles, creating a multi-target reflection setting \cite{ref:FLiu-ISAC}. Therefore, developing ISAC techniques that can effectively exploit this multi-target reflection property is essential for realizing robust system performance in real-world deployments.} 

Since localization has wide-ranging applications and also serves as the foundation to other sensing tasks such as tracking and imaging, in this paper we focus on how to embed a passive target localization functionality into 6G networks. 
Base station (BS) centric ISAC schemes where the target ``illumination'' is performed by one or more BSs (e.g., sending downlink beacon signals) and the sensing is performed by add-on specialized receivers, often co-located with BSs, have been widely investigated in the literature~\cite{ref:ISAC-Caire1,ref:ISAC-Caire2}. 
Note that a portion of time-frequency resource is reserved for uplink communication, and the signals from users can be scattered to receiving BSs via targets. It is thus promising to enhance the localization efficiency by utilizing uplink signals as well. However, the investigation of uplink ISAC is still in its infancy~\cite{ref:LLiu-ISAC}.

This paper highlights that massive communication, a key usage scenario in fifth-generation (5G) and 6G networks, designed to provide low-latency access to a large number of users, is well-suited for uplink localization, particularly in cell-free systems. In such systems, the uplink transmissions from a large number of distributed users provide extensive spatial coverage, enabling ubiquitous target localization. Moreover, the presence of multiple distributed BSs allows for cooperative processing of received echo signals, facilitating networked sensing through joint localization. Motivated by this, we present an investigation into the integration of massive communication and ISAC within the 6G cell-free framework. Specifically, in each time slot, a randomly selected subset of users becomes active, while multiple BSs collaboratively process the received signals to perform device activity detection, channel estimation, and target localization simultaneously.

\subsection{Related Works}

As to massive communication, or massive machine-type communications (mMTC) in 5G, due to the sparsity in user activity pattern, the joint user activity detection and channel estimation problem can be formulated as a compressed sensing problem and solved by a variety of algorithms including the low-complexity approximate message passing (AMP) \cite{ref:AMP} and its variants. 
In the scenario with unknown channel distribution, the denoiser of the AMP algorithm was designed under the minimax framework, leading to a soft-thresholding denoiser 
for the worst channel distribution~\cite{ref:AMP,ref:Zhu-learning}. 
In the scenario with known channel distribution, the minimum mean-squared error (MMSE) denoiser was designed via the Bayesian approach to minimize the mean-squared error~(MSE) for estimating the effective channels, where channel fading coefficients were assumed to follow independent and identically distributed (i.i.d.) Gaussian distributions~\cite{ref:Liu-massive}. 
Massive communication has also been studied via other variants of the AMP algorithm, including the generalized AMP~(GAMP)~\cite{ref:GAMP_propose, ref:Ke-EM, ref:Zhu1}, orthogonal AMP (OAMP)~\cite{ref:OAMP}, etc. 
The above works mainly focused on the single-BS scenario. 
An extension of the Bayesian AMP and OAMP for the cell-free scenario has been recently presented in \cite{ref:Caire-cellfreeAMP,ref:Caire-cellfreeAMP2}, under the name of ``multi source'' AMP and OAMP.

As to 6G localization, most of the works focus on downlink scenarios, i.e., the BSs emit wireless signals to localize the targets in the environment. Along this line, plenty of works have characterized the Cram\'er-Rao Bound~(CRB) of localization, and optimized the BS transmit beamforming vectors to reduce localization errors~\cite{ref:CRB1,ref:CRB2}. 
Besides the performance bound, other works have considered the algorithm design to localize the targets in practical 6G networks. For example, \cite{ref:downlink-MUSIC} proposed to utilize the multiple signal classification~(MUSIC) algorithm to localize targets. 
In a cell-free system, \cite{ref:QShi1} focused on networked sensing techniques, where multiple BSs share their echo signals to jointly localize the targets. 
Both theoretical performance guarantee and practical algorithm to deal with the data association issue were provided. 
{\blue The downlink ISAC model was considered in \cite{ref:downlink-ZhangRY1,ref:downlink-ZhangRY2}, where the BS simultaneously senses multiple targets and serves a communication user. A novel tensor-based approach was proposed to address both the channel estimation and target sensing problems.

In cellular networks, time and frequency resources are allocated to both uplink and downlink communication. Therefore, if the uplink signals can be utilized, the localization efficiency can be enhanced. This motivates us to explore the potential of integrating massive communication and target localization in this paper, where the uplink signals from active users can be utilized for localization.
In this scenario, the works \cite{ref:downlink-ZhangRY3,ref:Caire_ISAC,ref:ZhenGao} considered to localize targets while accomplishing the tasks of activity detection and channel estimation. In particular, all of these works 
adopted the sequential estimation method, where the compressed sensing algorithms were first utilized for activity detection and channel estimation, and then targets were localized based on the estimated channels.}  
However, considering that user activities, channels, and target positions are coupled together, such sequential approach fails to fully exploit the intricate dependency among them. Moreover, during the localization process, both the maximum \emph{a posterior} (MAP) estimation method in \cite{ref:Caire_ISAC} and the MUSIC-based scheme in \cite{ref:ZhenGao} require extensive searching operation to guarantee the performance, leading to prohibitive computational costs. Based on the above, it is necessary to explore how to jointly accomplish the massive communication and target localization tasks in an efficient manner for our considered ISAC system.

\subsection{Main Contributions}

\begin{figure}
    \centering
    \includegraphics[width=0.82\linewidth]{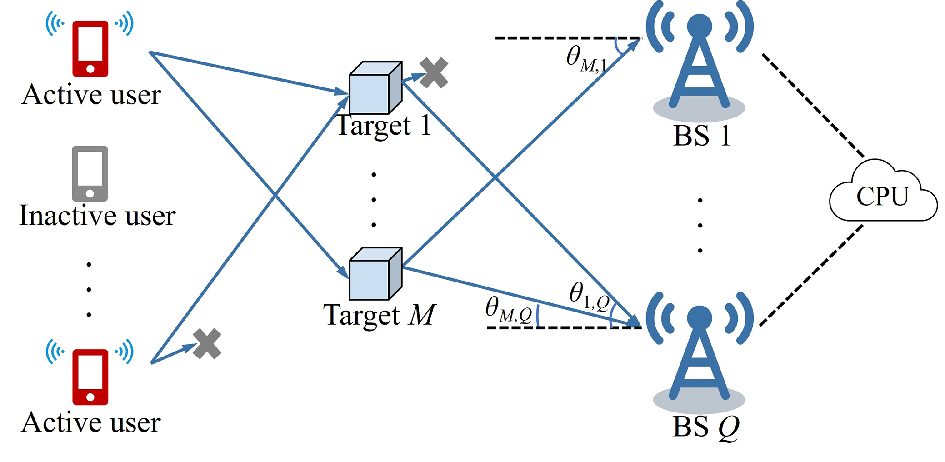}
    \caption{System model.}\vspace{-0.2cm}
    \label{fig:system_model}
\end{figure}

In this paper, we consider a 6G system consisting of a large number of users, multiple BSs, and multiple targets, as illustrated in Fig.~\ref{fig:system_model}. 
At each time slot, a random subset of users becomes active and transmits the pilot signals simultaneously, which are scattered by the targets and then received at the BSs. 
By aggregating the received signals at all BSs, the central processing unit (CPU) aims to accomplish two tasks: performing activity detection and channel estimation, and localizing the targets in the environment. 
The complex coupling between the unknown parameters poses significant challenges in achieving these tasks efficiently, which will be addressed in this work. 
The main contributions are summarized as follows.

\begin{itemize}

    \item  
    We build a probabilistic model to characterize the statistical relationship among user activities, channels, and target positions in the received signals at all BSs for the considered ISAC system. Upon this model, we propose to jointly estimate user activities, channels, and target locations under the MMSE criterion from the Bayesian estimation perspective.

    \item 
    We propose a novel turbo hybrid message passing (Turbo-HyMP) algorithm to solve the joint estimation problem, which is designed based on the factor graph of the probabilistic model. The proposed algorithm is comprised by two interconnected modules of target localization (TL) and activity detection and channel estimation (ADCE), which are iteratively executed to enhance the joint estimation performance. 
    Specifically, the TL module estimates target positions by extracting and fusing AoA information at each BS, 
    while the ADCE module addresses the joint estimation problem of user activities and channel coefficients. 
    The extrinsic messages of the estimated~AoAs and channels are exchanged between these two modules in each iteration, thereby refining the estimation
    in the next iteration. 
    By leveraging the turbo estimation procedure, the intricate dependency between channels and target positions can be fully exploited, which, however, is ignored in sequential estimation methods \cite{ref:downlink-ZhangRY3,ref:Caire_ISAC,ref:ZhenGao}.

    \item 
    Due to the complex coupling of variables, directly applying belief propagation (BP) based on the sum-product rule is computationally prohibitive. To overcome this challenge, we introduce approximate message computation techniques within the factor graph framework, enabling an efficient and tractable implementation. 
    Specifically, we approximate the messages related to AoAs and target positions with von Mises~(VM) distributions~\cite{ref:Statistics,ref:Badiu,ref:VM} and Gaussian distributions~\cite{ref:Kalman_mag,ref:Teng-Yuan}, respectively.
    Moreover, 
    GAMP is leveraged to simplify message passing in the dense bipartite sub-graph of the ADCE module. 
    These approximations result in a hybrid message passing structure, which balance 
    complexity and estimation accuracy while maintaining the advantages of~BP.

    \item 
    {\blue We provide Bayesian CRB (BCRB), state evolution (SE) analysis, complexity analysis, and extensive simulation results to verify the efficiency and effectiveness of the proposed algorithm.} 
    It is shown that the proposed scheme exhibits efficient convergence performance and consistent superiority over sequential counterparts. 
    Particularly, the proposed algorithm significantly enhances target localization performance with a reduction in localization error by about $6$ dB, as well as providing a reduction in the ADCE error by about $1.5$~dB. We interpret this performance gain as the ability of our scheme to efficiently exploit the statistical dependency among the unknown parameters. 

    
%

\end{itemize}


\emph{Organizations:}The rest of this paper is organized as follows. Section~\ref{sec:model} describes the ISAC system model. 
Section~\ref{Sec:formulation} formulates the considered problem. 
In Section~\ref{sec:TL-ADCE}, we establish a factor graph, based on which we devise the Turbo-HyMP algorithm. The BCRB, SE, and complexity analysis are provided in Section~\ref{Sec:performance}. 
Numerical results are presented in Section~\ref{Sec:simulation}. 
Finally, Section~\ref{ref:conclusion} concludes this paper.

  \emph{Notations:} Throughout this paper, uppercase and lowercase boldface letters denote matrices and vectors, respectively.
  The notation $\left[\mathbf{X} \right]_{m,:}$ denotes the $m$-th row of $\mathbf{X}$ and $\left[\mathbf{X} \right]_{:,n}$ denotes the $n$-th column of $\mathbf{X}$. 
  The notation $\mathbf{I}_{n}$ denotes an $n\times n$ identity matrix.
  We use $\left(\cdot \right)^{*}$, $\left(\cdot \right)^{T}$, $\left(\cdot \right)^{H}$, $\left\|\mathbf{x} \right\|_{p}$, $\left\|\mathbf{X} \right\|_{F}$, $\operatorname{tr}(\cdot)$, and $\odot$ to denote conjugate, transpose,
  conjugate transpose, ${\ell}_p$-norm, Frobenius norm, trace, and element-wise product, respectively. 
  We use $\operatorname{diag} \left\{ \mathbf{x} \right\}$ to denote a diagonal matrix with $\mathbf{x}$ comprising its diagonal elements, and use $\operatorname{blkdiag} \left\{ \mathbf{A}, \mathbf{B} \right\}$ to denote a block diagonal matrix. 
  Given any complex variable, vector, or matrix, we use $\Re(\cdot)$ and $\Im(\cdot)$ to return its real and imaginary parts, respectively.  
  We use $\cdot\backslash\cdot$ to denote set subtraction and $\left| \mathcal{A} \right|$ to denote the cardinality of a set $\mathcal{A}$. 
  {\blue For an integer $k > 0$, the notation $k!$ denotes the factorial function.} 
  The vectorization of a matrix is denoted as $\operatorname{vec}(\cdot)$.
  Let $[k] = \left\{1,\ldots,k \right\}$.
  We use $1[\cdot]$ to denote the indicator function and $\delta(\cdot)$ to denote the Dirac delta function.  
  $I_p(\cdot)$ denotes the modified Bessel function of the first kind and order $p$. 
  {\blue Let $f(x)$ and $g(x)$ be positive. The notation $f(x) = \mathcal{O}  \left( g(x)\right)$ means $\limsup_{x\to\infty} f(x)/g(x) < \infty$.}
  The distribution of a Bernoulli random variable $x$ with mean $\lambda$ is denoted as ${\rm{Ber}}(x;\lambda)$. 
  The distribution of a real Gaussian random vector $\mathbf{x}$ with mean $\bm{\mu}$ and covariance matrix $\bm{\Sigma}$ is denoted as $\mathcal{N}( \mathbf{x}; \bm{\mu} , \bm{\Sigma} )$. 
  The distribution of a circularly symmetric complex Gaussian~(CSCG) vector $\mathbf{x}$ with mean $\bm{\mu}$ and covariance matrix $\bm{\Sigma}$ is denoted as $\mathcal{CN}( \mathbf{x}; \bm{\mu} , \bm{\Sigma} )$. 
  We use $\mathcal{VM}(\theta;\eta)$ with $\eta = \kappa \exp(\jmath \mu)$ to denote the distribution of a VM random variable $\theta$, with its probability density function~(PDF) given~by 
\begin{subequations}
\begin{align} \label{eq:VMpdf1}
  \mathcal{VM} \left( \theta ; \eta \right) 
  & = \tfrac{1}{2\pi I_0\left( \left|\eta\right| \right)}  \exp\left( \Re\left\{ \eta^{*} \exp\left( \jmath \theta \right) \right\} \right) \\
  & = \tfrac{1}{2\pi I_0\left( \kappa \right)}  \exp\left( \kappa \cos \left( \theta - \mu \right)  \right). 
\end{align} 
\end{subequations}

\section{System Model}\label{sec:model}

We consider a cell-free ISAC system that integrates massive communication and localization as shown in Fig.~\ref{fig:system_model}. 
In this system, there are $K$ potential users ($K$ is large), $M$ targets, and $Q$ BSs. 
Each user is equipped with a single antenna, while each BS is equipped with a uniform linear array (ULA) consisting of $N$ antennas spaced by half of the carrier wavelength. BSs are connected to the CPU via fronthaul links. 
Under a two-dimensional 
Cartesian coordinate system, the locations of the $q$-th BS and the $m$-th target are denoted as $\mathbf{p}_{{\rm b},q} = \left[ x_{{\rm b},q},y_{{\rm b},q} \right]^T$ 
and $\mathbf{p}_{{\rm t},m} = \left[ x_{{\rm t},m} , y_{{\rm t},m} \right]^T$ 
in meter, respectively, $\forall q, m$. 
The locations of BSs are known, while those of targets are unknown. 
Similar to \cite{ref:Teng-Yuan,ref:GaussianP,ref:Kalman}, we model the location of target $m$ as a Gaussian random vector with mean $\bm{\mu}_{{\rm t},m}$ and covariance matrix $\mathbf{C}_{{\rm t},m}$, i.e., $\mathbf{p}_{{\rm t},m} \sim \mathcal{N} \left( \mathbf{p}_{{\rm t},m} ; \bm{\mu}_{{\rm t},m} , \mathbf{C}_{{\rm t},m}  \right) , \forall m$. 
In this work, we assume the distribution of $\mathbf{p}_{{\rm t},m}$ is known by exploiting historical data.\footnote{\blue 
One example with known target location distribution is tracking of moving targets. 
In this case, the target locations estimated in the previous frame give target location distribution in the current frame. The difference between the positions in any two adjacent frames can be modeled as an independent Gaussian noise, and thus Gaussian distribution is widely used to model the distribution of target positions \cite{ref:Kalman_mag,ref:Kalman,ref:Teng-Yuan}.}

At each coherence block, a random subset of the users will become active, and these active users transmit their pilot signals in the uplink simultaneously. 
These will reach the BSs through propagation channels that include the scattering due to the targets. 
In this context, the network (i.e., the ensemble of all BSs and the CPU) has two tasks: 
1) detecting the active users and estimating their channels; and 2) localizing the targets in the environment. 
Specifically, the pilot of user $k$ is denoted as $\mathbf{x}_{k} = [x_{k,1}, \ldots ,  x_{k,L} ]^T \in \mathbb{C}^{L \times 1}, \forall k$, where $x_{k,l}$ denotes the $l$-th pilot sample of user $k$, and $L$ denotes the length of the pilot signals. We assume that $x_{k,l}$'s are generated according to i.i.d. CSCG distribution with zero mean and variance $\frac{1}{L}$. 
Moreover, it is assumed that all the users decide in each coherence block whether or not to access the channel with a probability $\lambda$ in an i.i.d. manner. 
For convenience, we define a binary activity indicator $\alpha_k$ for each user $k$ as follows: 
\begin{equation} \label{eq:alphak}
 \alpha_k = \left\{ 
             \begin{array}{ll}
             1 , & \text{if user } k \text{ is active,} \\
             0 , & \text{otherwise.}
             \end{array} \right. 
\end{equation}
As a result, $\alpha_k\sim {\rm{Ber}}\left( \alpha_k ; \lambda \right), \forall k$. 
The signal received at the $q$-th BS is then given by
\begin{equation}\label{eq:receive_h}
    \mathbf{Y}_q = \sqrt{LP}\;\! {\sum}_{k=1}^{K} \alpha_{k} \mathbf{x}_k \mathbf{h}_{k,q}^T + \mathbf{Z}_q \in \mathbb{C}^{L\times N}, \;\;\; \forall q, 
\end{equation} 
where $P$ denotes the common transmit power of all the users, 
$\mathbf{Z}_q \in \mathbb{C}^{L \times N}$ denotes the noise with each element following i.i.d. CSCG distribution with zero mean and variance $\sigma_z^2$, 
and $\mathbf{h}_{k,q}$ denotes the channel from user $k$ to BS $q$. 
In practice, each user's signal is merely scattered by a subset of targets to each BS. 
In this paper, we use $s_{k,m,q}$'s to denote the indicators about presence of user-target-BS paths - $s_{k,m,q}$ is equal to one if the path from user $k$ to BS $q$ via target $m$ is not blocked, and zero otherwise. 
We assume that $s_{k,m,q} \sim \operatorname{Ber}(s_{k,m,q}; \xi), \forall k,m,q$, where $\xi$ denotes the common probability for the existence of a path. 
Then, the channel from user $k$ to BS $q$ can be modeled as
\begin{equation}\label{eq:model_h}
    \mathbf{h}_{k,q} = {\sum}_{m=1}^{M}  s_{k,m,q} 
    \rho_{k,m,q}  \mathbf{v}\! \left(\theta_{m,q}\right) \in \mathbb{C}^{N}, \;\;\; \forall k, q ,  
\end{equation}
where $\rho_{k,m,q}$ denotes the attenuation coefficient of this path contributed by path loss and radar cross section (RCS), 
$\theta_{m,q}$ denotes the AoA from the $m$-th target to the $q$-th BS given by
\begin{equation}\label{eq:sin_theta_target_BS}
  \theta_{m,q} = \arcsin  \left( \tfrac{y_{{\rm b},q} - y_{{\rm t},m}}{ d_{m,q} } \right) ,  \forall m,q,  
\end{equation} 
with $d_{m,q} = \left\| \mathbf{p}_{{\rm b},q} - \mathbf{p}_{{\rm t},m} \right\|_2 $, 
and $\mathbf{v}\left(\theta_{m,q}\right) = \left[ 1, e^{-j\pi\sin(\theta_{m,q})}, \ldots, e^{-j\pi(N-1)\sin(\theta_{m,q})} \right]^T \in \mathbb{C}^{N}$ denotes the receive steering vector of the BS towards $\theta_{m,q}$. 
Following the Swerling target model \cite{ref:RCS}, we assume that $\rho_{k,m,q}$'s follow the i.i.d. CSCG distribution with zero mean and variance $\sigma^2_{\rho,k,m,q}$, $\forall k,m,q$.\footnote{These hyperparameters in the system can be easily estimated based on the advanced machine learning techniques \cite{ref:Zhu-learning,ref:Ke-EM}.} 
For convenience, we define the effective attenuation coefficient associated with the path from user $k$ to target $m$ to BS $q$ as 
\begin{equation}\label{eq:b_kmq}
    b_{k,m,q} = \alpha_{k} s_{k,m,q} \rho_{k,m,q}, \;\; \forall k,m, q. 
\end{equation}
As a result, the effective channel attenuation of this path is equal to $\rho_{k,m,q}$ if and only if user $k$ is active, i.e., $\alpha_k=1$, and this link is not blocked, i.e., $s_{k,m,q}=1$.

Denote $\bar{\mathbf{X}} = \left[ \bar{\mathbf{x}}_1, \ldots,  \bar{\mathbf{x}}_K \right]$ with $\bar{\mathbf{x}}_k = \sqrt{LP} \mathbf{x}_k$, 
$\bm{\Lambda} = \operatorname{diag} \left\{ \alpha_1, \ldots, \alpha_K \right\}$, 
$\mathbf{V}_q = \left[ \mathbf{v} \! \left(\theta_{1,q}\right) , \ldots , \mathbf{v} \! \left(\theta_{M,q}\right) \right]^T$, 
$\bar{\mathbf{s}}_{m,q} = \left[ s_{1,1,q} \rho_{1,1,q}, \ldots, s_{K,1,q} \rho_{K,1,q} \right]^T$, and 
$\bar{\mathbf{S}}_q = \left[ \bar{\mathbf{s}}_{1,q} , \ldots , \bar{\mathbf{s}}_{M,q} \right]$. The received signal of BS $q$ given in \eqref{eq:receive_h} can be rewritten~as
\begin{equation}\label{eq:receive2}
    \mathbf{Y}_q 
    = \bar{\mathbf{X}} \bm{\Lambda} \bar{\mathbf{S}}_q \mathbf{V}_q  + \mathbf{Z}_q  , \;\;\; \forall q. 
\end{equation}   
The effective channel attenuation matrix of BS $q$ is defined as $\mathbf{B}_q = \bm{\Lambda} \bar{\mathbf{S}}_q \in \mathbb{C}^{K\times M}$, whose element on the $k$-th row and the $m$-th column is $b_{k,m,q}$ given in~\eqref{eq:b_kmq}. 
Denote the effective channel matrix $\bar{\mathbf{H}}_q = \mathbf{B}_q \mathbf{V}_q$. 
Each BS forwards its received signal to the CPU over the fronthaul link. 
{\blue In this paper, we assume that fronthaul transmission is perfect \cite{ref:Caire-cellfreeAMP, ref:downlink-ZhangRY3, ref:fronthaul1, ref:FLiu-ISAC2}. This is because we consider the pilot transmission phase, instead of the high-speed data transmission phase, and the current high capacity provided by fibers is sufficient for conveying pilot signals to the cloud.} 
Therefore, the CPU has the global and perfect received signals $\mathbf{Y} = \left[ \mathbf{Y}_1,  \ldots , \mathbf{Y}_Q\right]$. 

\section{Problem Statement}\label{Sec:formulation}



Based on the global information of $\mathbf{Y}$ at the CPU, our goal is to estimate three types of unknown variables - user activity indicators, i.e., $\alpha_k$'s given in \eqref{eq:alphak}, effective channel attenuation coefficients, i.e., $b_{k,m,q}$'s given in \eqref{eq:b_kmq}, and target locations, i.e., $\mathbf{p}_{{\rm t},m}$'s, which can be obtained after $\theta_{m,q}$'s are estimated based on the relation in \eqref{eq:sin_theta_target_BS}. 
Note that after estimating these variables, we can detect active users, estimate effective channels $\bar{\mathbf{H}}_q$'s, and localize targets.

The considered ISAC system integrates functionalities of massive communication and target localization. A straightforward approach is to sequentially estimate the communication-related and localization-related variables, starting from either the massive connectivity perspective or the localization perspective. 
On one hand, we can first perform joint activity detection and channel estimation by adopting compressed sensing techniques such as AMP/GAMP~\cite{ref:AMP, ref:Zhu-learning, ref:GAMP_propose, ref:Caire-cellfreeAMP}, and then extract AoAs from the targets to the BSs $\theta_{m,q}$'s from the estimated channels for target localization as in \cite{ref:ZhenGao}. 
On the other hand, we can also first leverage the standard AoA estimation schemes, such as the MUSIC algorithm~\cite{ref:MUSIC}, 
to estimate $\theta_{m,q}$'s, which are subsequently utilized to help solve the joint activity detection and channel estimation problem. 
However, the three types of unknown variables $\alpha_k$'s, $b_{k,m,q}$'s, and $\mathbf{p}_{{\rm t},m}$'s are coupled together. Therefore, such sequential approaches are generally suboptimal. 

Motivated by the above observation, in this work, we propose to exploit the implicit relation among $\alpha_k$'s, $b_{k,m,q}$'s, and $\mathbf{p}_{{\rm t},m}$'s for estimating them based on the global information of the received signals $\mathbf{Y}$ at the CPU, i.e., massive communication and target localization are performed simultaneously. 
Particularly, we will adopt the MMSE estimation technique. 
Given the global received signals $\mathbf{Y}$, the MMSE estimators of $\alpha_k$'s, $b_{k,m,q}$'s, and $\mathbf{p}_{{\rm t},m}$'s are respectively given as
\begin{align}
    \hat{\alpha}_k &= \int \alpha_k  p\left( \alpha_k | \mathbf{Y} \right)  {\rm d} \alpha_k , \;\; \forall k, \label{eq:hat_alpha} \\
    \hat{b}_{k,m,q} &= \int b_{k,m,q} p\left( b_{k,m,q} | \mathbf{Y} \right)  {\rm d} b_{k,m,q} , \;\; \forall k,m,q, \label{eq:hat_b} \\
    \hat{\mathbf{p}}_{{\rm t},m} &= \int \mathbf{p}_{{\rm t},m} p\left( \mathbf{p}_{{\rm t},m} | \mathbf{Y} \right)  {\rm d} \mathbf{p}_{{\rm t},m} , \;\; \forall m. \label{eq:hat_pt}
\end{align}
To obtain the MMSE estimators, we need to calculate the posterior marginal PDFs $p\left( \alpha_k | \mathbf{Y} \right)$, $p\left( \mathbf{p}_{{\rm t},m} | \mathbf{Y} \right)$, and $p\left( b_{k,m,q} | \mathbf{Y} \right), \forall k,m,q$. 
One straightforward way to calculate these posterior distributions is to first obtain the joint PDF of $\alpha_k$'s, $b_{k,m,q}$'s, $\mathbf{p}_{{\rm t},m}$'s, and $\mathbf{Y} , \forall k, m, q$, and then apply the Bayesian method to obtain the conditional marginal PDFs. However, because of the complicated relations between these variables, it is impossible to obtain the closed-form expression for such a joint PDF and the marginalization operation is generally computationally intractable.  
    
To tackle this issue, we introduce some intermediate variables to facilitate the calculation of the MMSE estimators. Specifically, let $\mathbf{C}_q = \bar{\mathbf{X}} \mathbf{B}_q \in \mathbb{C}^{L\times M}$ and $\mathbf{C} = [\mathbf{C}_1, \ldots, \mathbf{C}_Q]$. The $(l,m)$-th element of $\mathbf{C}_q$ is given by
\begin{equation}\label{eq:receive_target_c}
    c_{l,m,q} =  {\sum}_{k=1}^{K} b_{k,m,q} \bar{x}_{l,k} , \;\; \forall l, m, q,
\end{equation}  
where $\bar{x}_{l,k}$ denotes the $(l,k)$-th element of the matrix $\bar{\mathbf{X}}$. 
We define the $M\times Q$ matrix $\bm{\Delta}$ with $(m,q)$ element given by $\Delta_{m,q} = \pi \sin\left( \theta_{m,q} \right)$. 
Denote $\mathbf{B} = [\mathbf{B}_1, \ldots, \mathbf{B}_Q] \in \mathbb{C}^{K\times MQ}$ and $\mathbf{S} = [\mathbf{S}_1, \ldots, \mathbf{S}_Q] $ with the $(k,m)$-th element of $\mathbf{S}_q$ given by $s_{k,m,q}$. 
The relations between these variables are established as follows. 
First, following the geometric constraint in \eqref{eq:sin_theta_target_BS}, the conditional PDF $p\left( \left. \Delta_{m,q} \right| \mathbf{p}_{{\rm t},m} \right)$ is given by
\begin{equation}
    p\left( \left. \Delta_{m,q} \right| \mathbf{p}_{{\rm t},m} \right) = \delta \Big( \Delta_{m,q} 
    - \tfrac{\pi\left( y_{{\rm b},q} - y_{{\rm t},m} \right)}{ d_{m,q} } \Big) .
\end{equation} 
Second, because $\rho_{k,m,q}$ is CSCG distributed and $\alpha_k$ and $s_{k,m,q}$ are Bernoulli distributed, the conditional PDF $p\left( b_{k,m,q} \left| \alpha_k , s_{k,m,q} \right.  \right) $ is given as 
\begin{align} \label{eq:facnode_v}
    p\left( b_{k,m,q} \left| \alpha_k , s_{k,m,q} \right.  \right)  
    & = (1\!-\!\alpha_k s_{k,m,q}) \delta \left(   b_{k,m,q} \right)  \notag \\
    & \!\!\!\!\!  \!\!\!  + \alpha_k  s_{k,m,q} \mathcal{CN} \! \left( b_{k,m,q} ; 0 , \sigma^2_{\rho,k,m,q} \right) . 
\end{align} 
Next, according to \eqref{eq:receive_target_c}, the conditional PDF $p\left( \left. c_{l,m,q} \right| [\mathbf{B}_q]_{:,m}   \right)$ can be given as
\begin{equation} \label{eq:facnode_u}
    p\left( \left. c_{l,m,q} \right| [\mathbf{B}_q]_{:,m}   \right)  
    = \delta \left(  c_{l,m,q}  - [\bar{\mathbf{X}}]_{l,:} [\mathbf{B}_q]_{:,m}  
    \right) . 
\end{equation}
Denote the $n$-th element of $\mathbf{y}_{l,q}$ as $y_{l,n,q}$.  
Last, according to \eqref{eq:receive2}, the conditional PDF $p\left( \left. y_{l,n,q} \right| [\mathbf{C}_q]_{l,:} ,  [\bm{\Delta}]_{:,q} \right)$ is given as
\begin{align} \label{eq:facnode_g}
    & p\left( \left. y_{l,n,q} \right| [\mathbf{C}_q]_{l,:} ,  [\bm{\Delta}]_{:,q} \right) \notag \\ 
    & = \!\mathcal{CN} \! \left(  y_{l,n,q}  ; \! {\sum}_{m=1}^{M} \!c_{l,m,q} \!\exp\!\left( -\jmath (n\!-\!1) \Delta_{m,q} \right)  ,{\sigma}^2_{z} 
    \right) \!. 
\end{align}
Based on the above relation, the closed-form joint distribution of $p\left( \mathbf{p}_{{\rm t}} , \bm{\Lambda} , \mathbf{B} ,  \mathbf{S} , \mathbf{C} , \bm{\Delta} , \mathbf{Y} \right)$ is characterized as
\begin{align}
    & p\left( \mathbf{p}_{{\rm t}} , \bm{\Lambda} , \mathbf{B} ,  \mathbf{S} , \mathbf{C} , \bm{\Delta} , \mathbf{Y} \right) \notag \\
    & = \prod_{m=1}^{M} \! p\left( \mathbf{p}_{{\rm t},m} \right)  
    \prod_{k=1}^{K} \!p\left( \alpha_k \right)
    \prod_{k=1}^{K} \prod_{m=1}^{M} \prod_{q=1}^{Q} \!p\left( s_{k,m,q} \right) \notag\\
    & \;\;\;   \times 
    \prod_{k=1}^{K} \prod_{m=1}^{M} \prod_{q=1}^{Q} \!p\left( b_{k,m,q} \left| \alpha_k , s_{k,m,q} \right.  \right) 
    \!\prod_{m=1}^{M} \prod_{q=1}^{Q} \!p\left( \left. \Delta_{m,q} \right| \mathbf{p}_{{\rm t},m}  \right)   \notag\\ 
    & \;\;\;   \times 
    \prod_{l=1}^{L} \prod_{m=1}^{M} \prod_{q=1}^{Q}  p\left( \left. c_{l,m,q} \right| [\mathbf{B}_q]_{:,m} \right) 
    \notag\\
    & \;\;\;  \times
    \prod_{l=1}^{L} \prod_{q=1}^{Q} \prod_{n=1}^{N} p\left( \left. y_{l,n,q} \right| [\mathbf{C}_q]_{l,:}  ,  [\bm{\Delta}]_{:,q}  \right) , \label{eq:joint_PDF}
\end{align} 
where $p\!\left( \mathbf{p}_{{\rm t},m} \right) = \mathcal{N}\! \left( \mathbf{p}_{{\rm t},m} ; \bm{\mu}_{{\rm t},m} , \mathbf{C}_{{\rm t},m}  \right)$, $p\!\left( \alpha_k \right) \!=\! {\rm{Ber}}\left( \alpha_k ; \lambda \right)$, and $p\!\left( s_{k,m,q} \right) = {\rm{Ber}}\left( s_{k,m,q} ; \xi \right)$ denote prior distributions.

To summarize, although it is hard to obtain the joint PDF of $\alpha_k$'s, $b_{k,m,q}$'s, $\mathbf{p}_{{\rm t},m}$'s, and $\mathbf{Y} , \forall k, m, q$, we are able to characterize the joint PDF in \eqref{eq:joint_PDF} 
after introducing some intermediate variables. Then, 
following the Bayes' rule, the posterior marginal PDFs in \eqref{eq:hat_alpha}, \eqref{eq:hat_b}, and \eqref{eq:hat_pt} are derived by
\begin{equation} \label{eq:MAP_PDF_alphak}
    p\!\left( \alpha_k | \mathbf{Y} \right) 
    = \!\int\!  \tfrac{ p\left( \mathbf{p}_{{\rm t}} , \bm{\Lambda} , \mathbf{B} ,  \mathbf{S} , \mathbf{C} , \bm{\Delta} , \mathbf{Y} \right) }{ p\left( \mathbf{Y} \right) }  {\rm d} \mathbf{p}_{{\rm t}}  
    {\rm d} \bm{\Lambda}_{ \backslash \alpha_k } {\rm d} \mathbf{B} {\rm d} \mathbf{S} {\rm d} \mathbf{C} {\rm d} {\bm \Delta}, \;\;\!\!\!\forall k ,   
\end{equation} 
\begin{align} \label{eq:MAP_PDF_bkmq}
    p\!\left( b_{k,m,q} | \mathbf{Y} \right) 
    & = \int  \tfrac{ p\left( \mathbf{p}_{{\rm t}} , \bm{\Lambda} , \mathbf{B} ,  \mathbf{S} , \mathbf{C} , \bm{\Delta} , \mathbf{Y} \right) }{ p\left( \mathbf{Y} \right) }  \notag\\
    & \;\;\;\;\;  {\rm d} \mathbf{p}_{{\rm t}}  
    {\rm d} \bm{\Lambda} {\rm d} \mathbf{B}_{ \backslash b_{k,m,q} } {\rm d} \mathbf{S} {\rm d} \mathbf{C} {\rm d} {\bm \Delta}, \; \forall k,m,q ,
\end{align} 
\begin{align} \label{eq:MAP_PDF_ptm}
    p\!\left( \mathbf{p}_{{\rm t},m} | \mathbf{Y} \right) 
    & = \int  \tfrac{ p\left( \mathbf{p}_{{\rm t}} , \bm{\Lambda} , \mathbf{B} ,  \mathbf{S} , \mathbf{C} , \bm{\Delta} , \mathbf{Y} \right) }{ p\left( \mathbf{Y} \right) }   \notag\\
    & \;\;\;\;\; {\rm d} \mathbf{p}_{{\rm t} \backslash \mathbf{p}_{{\rm t},m} }  
    {\rm d} \bm{\Lambda} {\rm d} \mathbf{B} {\rm d} \mathbf{S} {\rm d} \mathbf{C} {\rm d} {\bm \Delta}, \;\; \forall m.
\end{align}  
Although the joint PDF is characterized, it is still computationally intractable to obtain the above posterior marginal PDFs due to high-dimensional integrals therein. In the rest of this paper, we apply the hybrid message passing algorithm to 
approximate these posterior marginal PDFs and thus approximate the MMSE estimators in \eqref{eq:hat_alpha}, \eqref{eq:hat_b}, and \eqref{eq:hat_pt}. 


\section{Joint Activity Detection, Channel Estimation, and Target Localization via Hybrid Message Passing}\label{sec:TL-ADCE}

This section introduces the proposed algorithm to solve the joint target localization, activity detection, and channel estimation problem. Specifically, we first establish a factor graph representation based on the probabilistic model introduced in Section~\ref{Sec:formulation}. 
Subsequently, a Turbo-HyMP algorithm is proposed to iteratively update the extrinsic messages passing along the edges in the factor graph for the approximations of the MMSE estimators in \eqref{eq:hat_alpha}, \eqref{eq:hat_b}, and \eqref{eq:hat_pt}. 

\subsection{Factor Graph Representation} \label{sec:factor_graph}

\begin{figure}
    \centering
    \includegraphics[width=0.998\linewidth]{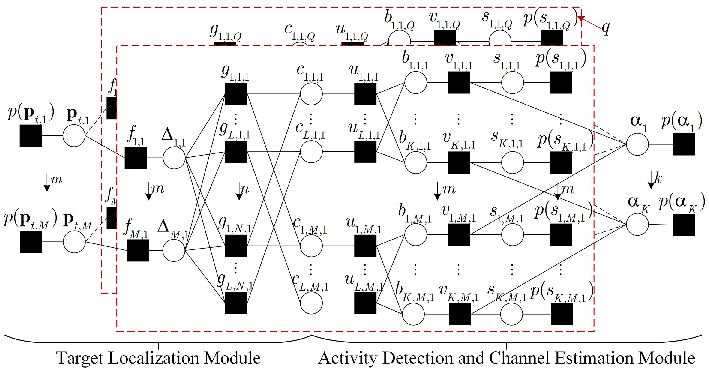}\\
    \caption{Factor graph representation of the joint distribution \eqref{eq:joint_PDF}.} \vspace{-0.2cm}
    \label{fig:factor_r}
\end{figure}

Upon the joint PDF in \eqref{eq:joint_PDF}, the factor graph representation can be given in Fig.~\ref{fig:factor_r}. 
Therein, blank circles represent variable nodes corresponding to the random variables $\{\mathbf{p}_{{\rm t},m}\}, \{\Delta_{m,q}\}, \{c_{l,m,q}\}, \{b_{k,m,q}\}, \{s_{k,m,q}\}$, and $\{\alpha_k\}$, 
while black rectangles represent factor nodes corresponding to functions $\{p\left( \left. \Delta_{m,q} \right| \mathbf{p}_{{\rm t},m} \right)\}$, 
$\{p\left( \left. y_{l,n,q} \right| [\mathbf{C}_q]_{l,:}  ,  [\bm{\Delta}]_{:,q}  \right)\}$, 
$\{p\left( \left. c_{l,m,q} \right| [\mathbf{B}_q]_{:,m} \right)\}$, 
and $\{p\left( b_{k,m,q} \left| \alpha_k , s_{k,m,q} \right.  \right)\}$, 
which are denoted as $\{f_{m,q}\}, \{g_{l,n,q}\}, \{u_{l,m,q}\}$, and $\{v_{k,m,q}\}$, respectively, for brevity. 
In particular, we propose to divide the factor graph into two modules: 
the TL and ADCE modules. 
As aforementioned, by introducing the auxiliary matrix $\mathbf{C}_q$, the received signal can be expressed as $\mathbf{Y}_q = \mathbf{C}_q \mathbf{V}_q + \mathbf{Z}_q$. 
With the extrinsic messages over $\mathbf{C}_q$ fed back from the ADCE module as the \textit{a priori}, the TL module is designed to extract AoA information from $\mathbf{Y}_q$, which is nontrivial due to the nonlinear dependency of $\mathbf{Y}_q$ on the angles. 
Then, the TL module fuses the AoA estimates at all BSs together to obtain the estimates of target locations, subsequently providing extrinsic messages over $\mathbf{C}_q$ for 
the ADCE module. 
{\blue Target localization relies on the sparsity in the angular domain of the channel since it is infeasible to estimate a significantly large number of AoAs from the received signal with limited dimension. As shown in Fig.~\ref{fig:factor_r}, to exploit this sparsity, our algorithm is conducted in the angular domain instead of the spatial domain, thereby reducing the number of parameters to be estimated.
} 
Moreover, the ADCE module aims to estimate user activities and effective attenuation coefficients based on the multi-user model $\mathbf{C}_q = \bar{\mathbf{X}} \bm{\Lambda} \bar{\mathbf{S}}_q$ and prior information of $\mathbf{C}_q$ from the TL module. 
By exchanging extrinsic messages between these two modules, both the TL and ADCE performance can be refined. 


\subsection{Turbo Hybrid Message Passing Algorithm}\label{Sec:Turbo_HyMP}

Based on the established factor graph, we develop a novel Turbo-HyMP algorithm to solve the joint activity detection, channel estimation, and target localization problem. 
Due to the complex coupling of variables, it is computationally prohibitive to directly apply the sum-product rule to update all the messages in the factor graph.
To address this issue, we propose to leverage the \emph{parameterized canonical distribution} method \cite{ref:MP-tem1,ref:MP-tem2} to simplify message calculations, which approximates PDFs in the propagated messages with a certain ``canonical'' family, characterized by some parameterization. 
Hence, we only need to update and propagate the PDF parameters, 
rather than the PDF itself. 
As a result, it is crucial to select suitable canonical distributions for the considered problem. 
Specifically, the AoA statistics are modeled with the VM distributions~\cite{ref:Badiu,ref:VM}, which are regarded as the circular analogues of normal distributions on the line \cite{ref:Statistics}. 
The statistics of target positions are modeled using Gaussian distributions~\cite{ref:Kalman_mag,ref:Teng-Yuan}, considering that their uncertainties usually result from the accumulation of small and independent errors. 
Thus, in this work, we approximate the messages exchanged between $\{f_{m,q}\}$ and $\{\Delta_{m,q}\}$ and those between $\{\Delta_{m,q}\}$ and $\{g_{l,n,q}\}$ using VM distributions, and approximate the messages from $\{\mathbf{p}_{{\rm t},m}\}$ to $\{f_{m,q}\}$ using Gaussian distributions.
With these approximations, the closed-form message updates of AoAs and positions are facilitated in the TL modules, thereby overcoming the intractable integrals.   
As a result, these approximations effectively balance computational complexity and estimation accuracy while maintaining the advantages of BP.  
Moreover, GAMP is adopted to simplify the message passing between $\{u_{l,m,q}\}$ and $\{b_{k,m,q}\}$ \cite{ref:GAMP_propose}. 
Due to its hybrid sparse-dense iterative message passing nature, the proposed algorithm is termed Turbo-HyMP.  
We outline the main steps of the proposed algorithm in Algorithm~\ref{alg:GAMP-MP}. 
{\blue For illustration, denote by $\mathcal{M}_{a\to b}^{(r)}(\cdot)$ the message from node $a$ to $b$ in the $r$-th iteration, and by $\mathcal{M}_{a}(\cdot)$ the marginal message at variable node $a$.} 
In the following, the details of all the message passing equations in the proposed Turbo-HyMP algorithm are introduced.


\begin{algorithm}[t] 
\caption{Turbo-HyMP}
\label{alg:GAMP-MP} 
\begin{algorithmic}[1] 
{ \small
\REQUIRE 
Pilot $\mathbf{X}$;
received signal $\mathbf{Y}$; 
mean $\bm{\mu}_{{\rm t},m}$ and variance $\mathbf{C}_{{\rm t},m}$ of the position of target $m, \forall m$; 
position $\mathbf{p}_{{\rm b},q}$ of BS $q, \forall q$; 
noise variance $\sigma_{z}^2$;
tolerance $\epsilon_{p}$; 
and iterations ${\text{Iter}_{\rm out}}$~and~${\text{Iter}_{\max}}$.

\ENSURE 
$\hat{\mathbf{p}}_{{\rm t},m}, \hat{\alpha}_k$, and $\hat{\mathbf{h}}_{k,q}$.


\STATE \textbf{Initialize:} 
$\forall l,m,n,q,k:$ $\eta_{ f_{m,q} \to  \mathbf{p}_{{\rm t},m} }^{(1,1)}= 0$, $\eta_{ g_{l,n,q} \to  \Delta_{m,q}  }^{(1,1)}= 0$, $\mu_{  c_{l,m,q}  \to  g_{l,n,q} }^{(1,1)}= 0$, $(\sigma^{(1,1)}_{  c_{l,m,q}  \to  g_{l,n,q} })^2 = 10^6$, $\mu_{   u_{l,m,q} \to c_{l,m,q}  }^{(1)} = 0 , (\sigma^{(1)}_{   u_{l,m,q} \to c_{l,m,q}  })^2 = 10^6$,
and $\lambda_{ v_{k,m,q} \to \alpha_k }^{(1)}= 0.5$.\\

\FOR{$i = 1,2,\ldots, {\text{Iter}_{\rm out}}$} 

\FOR{$j = 1,2,\ldots, {\text{Iter}_{\max}}$} 

\STATE \textbf{\% TL module (forward)}
\STATE $\forall m, q:$ Update $\bm{\mu}_{\mathbf{p}_{{\rm t},m} \to f_{m,q}}^{(i,j)}$ and $\mathbf{C}_{\mathbf{p}_{{\rm t},m} \to f_{m,q}}^{(i,j)}$ via Newton-Gauss method, \eqref{eq:message_I_xy_fq_mu}, and \eqref{eq:message_I_xy_fq_var}. 

\STATE $\forall m, q:$ Update $\eta_{f_{m,q} \to \Delta_{m,q}}^{(i,j)}$ via \eqref{eq:message_I_fq_delta_eta}.

\STATE $\forall l,n,m, q:$ Update $\eta_{ \Delta_{m,q} \to g_{l,n,q} }^{(i,j)} $ by \eqref{eq:message_I_delta_g_eta} and \cite[Alg.~2]{ref:Badiu}.

\STATE $\forall l,n,m, q:$ Update $\mu_{ g_{l,n,q} \to c_{l,m,q} }^{(i,j)}$ and $\big(\sigma^{(i,j)}_{ g_{l,n,q} \to c_{l,m,q} }\big)^2$ via \eqref{eq:message_I_g_c_mu} and \eqref{eq:message_I_g_c_sigma}.

\STATE \textbf{\% TL module (backward)}

\STATE $\forall l,n,m, q:$ Update $\mu_{  c_{l,m,q}  \to  g_{l,n,q} }^{(i,j+1)}$ and $\big(\sigma^{(i,j+1)}_{  c_{l,m,q}  \to  g_{l,n,q} }\big)^2$ via \eqref{eq:message_I_c_g_mu} and \eqref{eq:message_I_c_g_sigma}.

\STATE $\forall l,n,m, q:$ Update $\eta_{ g_{l,n,q} \to  \Delta_{m,q}  }^{(i,j+1)}$ via \eqref{eq:message_I_g_delta_eta}.

\STATE $\forall m, q:$ Update $\eta_{ \Delta_{m,q} \to f_{m,q} }^{(i,j+1)}$ applying \cite[Alg.~2]{ref:Badiu}.

\STATE $\forall m, q:$ Update $\eta_{ f_{m,q} \to  \mathbf{p}_{{\rm t},m} }^{(i,j+1)}$ via \eqref{eq:message_I_fq_pt_eta}.

\ENDFOR

\STATE \textbf{\% ADCE module (into)} 

\STATE $\forall l,m, q:$ Update $\mu_{ c_{l,m,q}  \to u_{l,m,q} }^{(i)}$ and $\big(\sigma^{(i)}_{ c_{l,m,q}  \to u_{l,m,q} }\big)^2$ via \eqref{eq:message_I_c_u_mu} and \eqref{eq:message_I_c_u_sigma}.

\STATE $\forall k,m, q:$ Update $\lambda_{ \alpha_k \to v_{k,m,q} }^{(i)}$ via \eqref{eq:message_I_alpha_v_lambda}. 

\STATE $\forall k,m, q:$ Update $\xi_{ v_{k,m,q} \to b_{k,m,q} }^{(i)}$ via \eqref{eq:message_I_v_b_xi}.

\STATE \textbf{\% ADCE module (within)}

\STATE Run the GAMP algorithm \cite{ref:GAMP_propose}.

\STATE \textbf{\% ADCE module (out)}  

\STATE $\forall k,l,m, q:$ Update \eqref{eq:message_I_u_c} and \eqref{eq:message_I_b_v}.

\STATE $\forall k,m, q:$ Update $\lambda_{ v_{k,m,q} \to \alpha_k }^{(i+1)}$ via \eqref{eq:message_I_vkmq_alphak_lambda}.



\STATE {$\forall m:$ Update $\!\hat{\bm{\mu}}_{{\rm t},m}^{\!(i\!+\!1,  {\rm Iter}_{\max})}\!$ via $\!\mathcal{M}_{ f_{m,q} \to  \mathbf{p}_{{\rm t},m} }^{(i+1, {\rm Iter}_{\max})} \!\! \left( \mathbf{p}_{{\rm t},m} \right)\!$ as in~\eqref{eq:message_I_xy_output_mu}.}

\STATE If $ \sqrt{\sum_m \big\| \hat{\bm{\mu}}_{{\rm t},m}^{(i+1, {\rm Iter}_{\max})} - \hat{\bm{\mu}}_{{\rm t},m}^{(i, {\rm Iter}_{\max})}  \big\|_2^2 / M  }\leq \epsilon_{p}$, stop.

\ENDFOR

\STATE $\forall k,m, q:$ Update $\hat{\mathbf{p}}_{{\rm t},m}, \hat{\alpha}_k$, and $\hat{\mathbf{h}}_{k,q}$ via \eqref{eq:message_I_xy_output_ptm}, \eqref{eq:output_alpha}, and \eqref{eq:output_h}.

}

\end{algorithmic}
\end{algorithm}

\textbf{TL Module:} 
{\blue Based on the received signal $\mathbf{Y}_q = \mathbf{C}_q \mathbf{V}_q + \mathbf{Z}_q$, the extrinsic messages over $\mathbf{C}_q$ from the ADCE module, and the prior information on target positions, the TL module is designed to update AoA and target position estimates and provide extrinsic messages over $\mathbf{C}_q$ for the subsequent message update in the ADCE module.  
The TL module operates in two phases: the (forward)-phase and the (backward)-phase. 
In the (forward)-phase, the message is calculated through the path $\mathbf{p}_{{\rm t},m} \to f_{m,q} \to \Delta_{m,q} \to g_{l,n,q} \to c_{l,m,q}$, in which target positions are used to update the estimates of AoAs $\Delta_{m,q}$'s and the auxiliary variables $c_{l,m,q}$'s (see Lines 5-8 of Algorithm~\ref{alg:GAMP-MP}). 
With the messages over $c_{l,m,q}$ from the ADCE module, the backward message is calculated through the path $c_{l,m,q} \to g_{l,n,q} \to \Delta_{m,q} \to f_{m,q} \to \mathbf{p}_{{\rm t},m}$ to update AoA and target position information (see Lines 10-13 of Algorithm~\ref{alg:GAMP-MP}).} 
In the $i$-th outer iteration, we perform message passing within the TL module for several iterations for stability. 
The messages 
in the $j$-th inner iteration are described sequentially as follows.

\subsubsection{Message from $\mathbf{p}_{{\rm t},m}$ to $f_{m,q}$}
The message from the variable node $\mathbf{p}_{{\rm t},m}$ to the factor node $f_{m,q}$ is given by
\begin{subequations}
\begin{align}
  & \mathcal{M}_{  \mathbf{p}_{{\rm t},m} \to f_{m,q} }^{(i,j)} \left( \mathbf{p}_{{\rm t},m} \right) \notag\\
  & = p\left( \mathbf{p}_{{\rm t},m} \right) 
  {\prod}_{q'\neq q} \mathcal{M}_{ f_{m,q'} \to  \mathbf{p}_{{\rm t},m} }^{(i,j)}  \left( \mathbf{p}_{{\rm t},m} \right)  \label{eq:message_I_xy_fq_prod} \\
  & \approx \mathcal{N} \left( \mathbf{p}_{{\rm t},m} ; \bm{\mu}_{{\rm t},m} , \mathbf{C}_{{\rm t},m} \right) 
    \mathcal{N} \left( \mathbf{p}_{{\rm t},m} ; {\tilde{\bm{\mu}}}_{{\rm t},m,q}^{(i,j)}  , {\tilde{\mathbf{C}}}_{{\rm t},m,q}^{(i,j)}   \right) \label{eq:message_I_xy_fq_approG} \\
  & \propto \mathcal{N} \left( \mathbf{p}_{{\rm t},m} ; \bm{\mu}_{\mathbf{p}_{{\rm t},m} \to f_{m,q}}^{(i,j)} , \mathbf{C}_{\mathbf{p}_{{\rm t},m} \to f_{m,q}}^{(i,j)} \right) .  \label{eq:message_I_xy_fq}
\end{align}
\end{subequations}    
{\blue Here, \eqref{eq:message_I_xy_fq_approG} follows by approximating the message 
$\prod_{q'\neq q}  \mathcal{M}_{ f_{m,q'} \to \mathbf{p}_{{\rm t},m} }^{(i,j)} \left( \mathbf{p}_{{\rm t},m} \right)$ as a Gaussian PDF applying the second-order Taylor approximation, where the expansion point is a local maximizer of $\prod_{q'\neq q}  \mathcal{M}_{ f_{m,q'} \to \mathbf{p}_{{\rm t},m} }^{(i,j)} \left( \mathbf{p}_{{\rm t},m} \right)$ obtained by Newton-Gauss iterations~\cite{ref:Newton_Gauss}. 
Detailed derivation of ${\tilde{\bm{\mu}}}_{{\rm t},m,q}^{(i,j)}$ and ${\tilde{\mathbf{C}}}_{{\rm t},m,q}^{(i,j)}$ in \eqref{eq:message_I_xy_fq_approG} is provided in Appendix~\ref{Appendix_proof_I_xy_fq}.} 
\eqref{eq:message_I_xy_fq} follows from Gaussian multiplication rules with mean and variance given by
\begin{subequations}
\begin{align}
  & \bm{\mu}_{\mathbf{p}_{{\rm t},m} \to f_{m,q}}^{(i,j)}  
  = \left[ \mu_{\mathbf{p}_{{\rm t},m} \to f_{m,q} , x}^{(i,j)}  , \mu_{\mathbf{p}_{{\rm t},m} \to f_{m,q} , y}^{(i,j)} \right]^T \\
  & = \mathbf{C}_{\mathbf{p}_{{\rm t},m} \to f_{m,q}}^{(i,j)} 
  \left( {\mathbf{C}}_{{\rm t},m}^{-1} \bm{\mu}_{{\rm t},m}  + \big({\tilde{\mathbf{C}}}_{{\rm t},m,q}^{(i,j)}\big)^{-1}  {\tilde{\bm{\mu}}}_{{\rm t},m,q}^{(i,j)} \right),  \label{eq:message_I_xy_fq_mu}
\end{align}
\end{subequations}
\begin{equation}
  \mathbf{C}_{\mathbf{p}_{{\rm t},m} \to f_{m,q}}^{(i,j)}
  = \left(  {\mathbf{C}}_{{\rm t},m}^{-1} + \big({\tilde{\mathbf{C}}}_{{\rm t},m,q}^{(i,j)}\big)^{-1} \right)^{-1}. \label{eq:message_I_xy_fq_var}
\end{equation}

\subsubsection{Message from $f_{m,q}$ to $\Delta_{m,q}$} 
The message from factor node $f_{m,q}$ to variable node $\Delta_{m,q}$ is expressed as follows:
\begin{align}
  & \mathcal{M}_{  f_{m,q} \to \Delta_{m,q} }^{(i,j)}   \left( \Delta_{m,q} \right) \notag\\ 
  & =   \int   \mathcal{M}_{ \mathbf{p}_{{\rm t},m} \to f_{m,q} }^{(i,j)}  \left( \mathbf{p}_{{\rm t},m} \right)   p \left(  \left. \Delta_{m,q} \right| \mathbf{p}_{{\rm t},m} \right)  {\rm d}\mathbf{p}_{{\rm t},m}  \label{eq:message_I_fq_delta0} .  
\end{align}
 
The integral in \eqref{eq:message_I_fq_delta0} does not have a closed-form expression. To facilitate tractable message passing, we follow the idea in \cite[Appendix~B]{ref:Teng-Yuan} 
to approximate this message by a VM PDF. 
Specifically, we denote the unit directional vector perpendicular to $\bm{\mu}_{\mathbf{p}_{{\rm t},m} \to f_{m,q}}^{(i,j)} - \mathbf{p}_{{\rm b},q}$ as $\mathbf{v}_{m,q}^{(i,j)}$. We represent the projection of $\mathbf{p}_{{\rm t},m} - \bm{\mu}_{\mathbf{p}_{{\rm t},m} \to f_{m,q}}^{(i,j)}$ onto $\mathbf{v}_{m,q}^{(i,j)}$ as a random variable $\zeta_{m,q}$ satisfying $\zeta_{m,q} \!\sim\! \mathcal{N} \!\left( \zeta_{m,q} ; 0 , \big(\mathbf{v}_{m,q}^{(i,j)}\big)^{\!T} \!  \mathbf{C}_{\mathbf{p}_{{\rm t},m} \to f_{m,q}}^{(i,j)} \mathbf{v}_{m,q}^{(i,j)} \right)$. 
In the far-field environment, this reduces the integral in \eqref{eq:message_I_fq_delta0} over the vector $\mathbf{p}_{{\rm t},m}$ to an integral over the variable $\zeta_{m,q}$, which can be easily calculated. 
Then, applying the second-order Taylor expansion, we approximate \eqref{eq:message_I_fq_delta0} by a VM PDF as follows:   
\begin{equation}
    \mathcal{M}_{ f_{m,q} \to \Delta_{m,q} }^{(i,j)} \left( \Delta_{m,q} \right)
    \approx \mathcal{VM} \left(  \Delta_{m,q} ; \eta_{f_{m,q} \to \Delta_{m,q}}^{(i,j)} \right) , \label{eq:message_I_fq_delta}
\end{equation}  
where 
\begin{equation} \label{eq:message_I_fq_delta_eta}
  \eta_{ f_{m,q} \to \Delta_{m,q} }^{(i,j)} = \kappa_{ f_{m,q} \to \Delta_{m,q} }^{(i,j)}  \exp\left( \jmath \mu_{ f_{m,q} \to \Delta_{m,q} }^{(i,j)} \right), 
\end{equation}
\begin{equation}
    \mu_{ f_{m,q} \to \Delta_{m,q} }^{(i,j)} 
    = \tfrac{\pi\left( y_{{\rm b},q} - \mu_{\mathbf{p}_{{\rm t},m} \to f_{m,q} , y}^{(i,j)} \right)}{  \left\| {\mathbf{p}}_{{\rm b},q} - \bm{\mu}_{\mathbf{p}_{{\rm t},m} \to f_{m,q} }^{(i,j)}  \right\|_2  }  ,\label{eq:message_I_fq_delta_mu}
\end{equation}
\begin{equation}
    \kappa_{f_{m,q} \to \Delta_{m,q}}^{(i,j)}  = \tfrac{ \left\| {\mathbf{p}}_{{\rm b},q} - \bm{\mu}_{\mathbf{p}_{{\rm t},m} \to f_{m,q} }^{(i,j)}  \right\|_2^2  }{ \!\left( \pi^2 - \left(\mu_{ f_{m,q} \to \Delta_{m,q} }^{(i,j)}\right)^{\!2} \right) 
    \!\left(\mathbf{v}_{m,q}^{(i,j)}\right)^{\!T}\!  \mathbf{C}_{\mathbf{p}_{{\rm t},m} \to f_{m,q}}^{(i,j)} \!\mathbf{v}_{m,q}^{(i,j)} } . \label{eq:message_I_fq_delta_kappa}
\end{equation}

\subsubsection{Message from $\Delta_{m,q}$ to $g_{l,n,q}$} The message from variable node $\Delta_{m,q}$ to factor node $g_{l,n,q}$ is expressed as
\begin{align}
  & \mathcal{M}_{ \Delta_{m,q} \to g_{l,n,q} }^{(i,j)} \left( \Delta_{m,q} \right) =  \mathcal{M}_{ f_{m,q} \to \Delta_{m,q} }^{(i,j)} \left( \Delta_{m,q} \right)  \notag\\
  & \;\;\;\;\; \;\;\;\;\; \;\;\;\;\; \;\;\;\;\;   \times {\prod}_{(l',n')\neq (l,n)}  \mathcal{M}_{ g_{l',n',q} \to \Delta_{m,q} }^{(i,j)} \left( \Delta_{m,q} \right) . \label{eq:message_I_delta_g0} 
\end{align}

The second term on the RHS of \eqref{eq:message_I_delta_g0} satisfies
\begin{subequations}
\begin{align}
  & {\prod}_{(l',n')\neq (l,n)}  \mathcal{M}_{ g_{l',n',q} \to \Delta_{m,q} }^{(i,j)} \left( \Delta_{m,q} \right) \notag \\
  & \propto {\prod}_{(l',n')\neq (l,n)}  \mathcal{VM} \left(  (n'-1) \Delta_{m,q} ; \eta_{g_{l',n',q} \to \Delta_{m,q}}^{(i,j)} \right) \label{eq:message_I_delta_g_VMnl} \\ 
  & \propto {\prod}_{ n'\geq 2 }  \mathcal{VM} \left(   (n'-1) \Delta_{m,q} ; \eta_{\Delta,l,m,q,n,n'}^{(i,j)}  \right) \label{eq:message_I_delta_g_VMn} \\
  & \approx  \mathcal{VM} \!\left(  \Delta_{m,q} ; \bar{\eta}_{\Delta,l,m,q,n}^{(i,j)} \right) , \label{eq:message_I_delta_g_VM}
\end{align}
\end{subequations}
{\blue where \eqref{eq:message_I_delta_g_VMnl} holds because we approximate the message from $g_{l',n',q}$ to $\Delta_{m,q}$ as a $(n'-1)$-fold wrapped VM PDF, as will be introduced in \eqref{eq:message_I_g_delta}; 
\eqref{eq:message_I_delta_g_VMn} holds because the family of VM PDFs is closed under multiplication \cite[Eq. (27)]{ref:Badiu} with $\eta_{\Delta,l,m,q,n,n'}^{(i,j)} = \kappa_{\Delta,l,m,q,n,n'}^{(i,j)} \exp\left( \jmath \mu_{\Delta,l,m,q,n,n'}^{(i,j)} \right) $ given by 
\begin{equation}
 {\mathop{\eta}}_{\Delta,l,m,q,n,n'}^{(i,j)} = \left\{ 
             \begin{array}{lr}
             \sum_{l'} \eta_{g_{l',n',q} \to \Delta_{m,q}}^{(i,j)} , & \! n' \neq n, \\
             \sum_{l'\neq l} \eta_{g_{l',n',q} \to \Delta_{m,q}}^{(i,j)} , & \! n' = n.
             \end{array} \right.  \label{eq:message_I_delta_g_VM_eta}
\end{equation}  
To obtain \eqref{eq:message_I_delta_g_VM}, we first approximate the $(n'-1)$-fold wrapped VM PDF in \eqref{eq:message_I_delta_g_VMn} as a mixture of $n'-1$ VM PDFs based on \cite[Eq. (30)]{ref:Badiu}. 
However, it leads to an intractable number $(N-1)!$ of components. 
To this end, we adopt the heuristic method in \cite[Algorithm~2]{ref:Badiu} to represent \eqref{eq:message_I_delta_g_VMn} by a single VM PDF, in which we first search for the most dominant component and then represent \eqref{eq:message_I_delta_g_VMn} by a single VM PDF based on a second-order Taylor approximation around the mean of this dominant component. 
}


Substituting \eqref{eq:message_I_delta_g_VM} into \eqref{eq:message_I_delta_g0}, we have
\begin{subequations}
\begin{align}
  & \mathcal{M}_{ \Delta_{m,q} \to g_{l,n,q} }^{(i,j)} \left( \Delta_{m,q} \right) \notag\\
  & \approx \mathcal{VM} \left(  \Delta_{m,q} ; \eta_{f_{m,q} \to \Delta_{m,q}}^{(i,j)} \right) 
  \mathcal{VM} \!\left(  \Delta_{m,q} ; \bar{\eta}_{\Delta,l,m,q,n}^{(i,j)} \right) \\
  & \propto \mathcal{VM} \left( \Delta_{m,q} ; \eta_{ \Delta_{m,q} \to g_{l,n,q} }^{(i,j)} \right) , \label{eq:message_I_delta_g}
\end{align} 
\end{subequations}
where \eqref{eq:message_I_delta_g} holds because the family of VM PDFs is closed under multiplication \cite[Eq. (27)]{ref:Badiu} with $\eta_{ \Delta_{m,q} \to g_{l,n,q} }^{(i,j)}$ given by
\begin{subequations}
\begin{align}
    \eta_{ \Delta_{m,q} \to g_{l,n,q} }^{(i,j)} &=   \kappa_{ \Delta_{m,q} \to g_{l,n,q} }^{(i,j)} \!\exp\!\left( \jmath \mu_{ \Delta_{m,q} \to g_{l,n,q} }^{(i,j)} \right) \\
    & = \eta_{f_{m,q} \to \Delta_{m,q}}^{(i,j)} + \bar{\eta}_{\Delta,l,m,q,n}^{(i,j)}. \label{eq:message_I_delta_g_eta}
\end{align}  
\end{subequations}

\subsubsection{Message from $g_{l,n,q}$ to $c_{l,m,q}$} The message from factor node $g_{l,n,q}$ to variable node $c_{l,m,q}$ is given by 
\begin{subequations}
\begin{align}
  & \mathcal{M}_{ g_{l,n,q} \to c_{l,m,q} }^{(i,j)} \left(c_{l,m,q} \right) \notag\\
  & = \!\int \!\! \prod_{m' = 1}^{M} \! \mathcal{M}_{ \Delta_{m',q} \to g_{l,q,n} }^{(i,j)} \!\!\left( \Delta_{m',q} \right)  
  \!\!\prod_{m'\neq m} \!\!\mathcal{M}_{  c_{l,m',q}  \to  g_{l,n,q} }^{(i,j)} \!\!\left( c_{l,m',q} \right)   \notag\\
  & \;\;\;\; \times p\!\left( \left. y_{l,n,q} \right| [\mathbf{C}_q]_{l,:}  ,  [\bm{\Delta}]_{:,q}  \right) {\rm d} [\mathbf{C}_q]_{l,:  \backslash c_{l,m,q} }  {\rm d}   [\bm{\Delta}]_{:,q   }   \label{eq:message_I_g_c_int}  \\
  & = \!\int \!\! \mathcal{M}_{ \Delta_{m,q} \to g_{l,q,n} }^{(i,j)} \!\!\left( \Delta_{m,q} \right)  
  f_n^{(i,j)} \!\left( c_{l,m,q} , \Delta_{m,q} \right) {\rm d} \Delta_{m,q} ,  \label{eq:message_I_g_c_int2} 
\end{align}
\end{subequations}
where $f_n^{(i,j)} \left( c_{l,m,q} , \Delta_{m,q} \right)$ is given by
\begin{align}
  & f_n^{(i,j)} \left( c_{l,m,q} , \Delta_{m,q} \right)  \notag\\
  & = \!\!\int\!\! \prod_{m'\!\neq m} \!\!\mathcal{M}_{  c_{l,m',q}  \to  g_{l,n,q} }^{(i,j)} \!\!\left( c_{l,m',q} \right)  \! \prod_{m'\!\neq m} \!\!\mathcal{M}_{ \Delta_{m',q} \to g_{l,n,q} }^{(i,j)} \!\!\left( \Delta_{m',q} \right) 
  \notag\\   
  & \;\;  \times \!p\!\left( \left. y_{l,n,q} \right| [\mathbf{C}_q]_{l,:}  ,  [\bm{\Delta}]_{:,q}  \right) {\rm d} [\mathbf{C}_q]_{l,: \backslash c_{l,m,q}}  {\rm d}  [\bm{\Delta}]_{:,q \backslash \Delta_{m,q} }  .  \label{eq:message_I_g_delta_int_part2}
\end{align}  
The high-dimensional integrals in \eqref{eq:message_I_g_delta_int_part2} lead to an unacceptably high complexity. 
To address this issue, we employ the moment matching principle and the characteristic function of VM distributions~\cite{ref:Statistics} to approximate 
\eqref{eq:message_I_g_delta_int_part2} by a Gaussian PDF
\begin{equation}  
    f_n^{(i,j)} \!  \left( c_{l,m,q} , \Delta_{m,q} \right)
    \!\approx \!
    \mathcal{CN} \!\!\left(\! c_{l,m,q}  ; \mu_{l,m,q,n}^{(i,j)} , \big(\sigma^{(i,j)}_{l,m,q,n}\big)^{\!2}  \right) \! .  \label{eq:message_I_g_c_int_part2_G}
\end{equation}
The mean $\mu^{(i,j)}_{l,m,q,n}$ and variance $\big( \sigma^{(i,j)}_{l,m,q,n} \big)^2$ are given by 
\begin{equation}
    \mu_{l,m,q,n}^{(i,j)}
    = \bar{\mu}_{l,m,q,n}^{(i,j)}  \exp\left( \jmath (n-1) \Delta_{m,q} \right), \label{eq:message_I_g_delta_int_part2_G_mu}
\end{equation}
\begin{align}
  & \big(\sigma^{(i,j)}_{l,m,q,n}\big)^2  =  \sigma^2_z  
  + \!\sum_{m'\neq m} \! \big(\sigma^{(i,j)}_{  c_{l,m',q}  \to  g_{l,n,q} }  \big)^2
  \notag\\
  & \;\;\;\;  + \! \sum_{m'\neq m} \!\! \big| \mu_{  c_{l,m',q}  \to  g_{l,n,q} }^{(i,j)} \big|^2 
  \Bigg( 1 \! - \!
  \tfrac{I^2_{n-1} \! \big( \kappa_{ \Delta_{m',q} \to g_{l,n,q} }^{(i,j)} \big)}{I^2_0\big( \kappa_{ \Delta_{m',q} \to g_{l,n,q} }^{(i,j)} \big)}  \Bigg) ,\label{eq:message_I_g_c_int_part2_G_sigma2}
\end{align} 
\begin{align}
  \bar{\mu}_{l,m,q,n}^{(i,j)}  
  & = y_{l,n,q} - {\sum}_{m'\neq m} \!  \exp\!\left( - \jmath (n-1) \mu_{ \Delta_{m',q} \to g_{l,n,q} }^{(i,j)} \right)  \notag\\
  & \;\;\;\;\; \times \mu_{  c_{l,m',q}  \to  g_{l,n,q} }^{(i,j)}  \tfrac{I_{n-1} \big( \kappa_{ \Delta_{m',q} \to g_{l,n,q} }^{(i,j)} \big)}{I_0\big( \kappa_{ \Delta_{m',q} \to g_{l,n,q} }^{(i,j)} \big)} . \label{eq:message_I_g_c_int_part2_G_barmu}
\end{align} 

Substituting \eqref{eq:message_I_delta_g} and \eqref{eq:message_I_g_c_int_part2_G} into \eqref{eq:message_I_g_c_int2}, we have
\begin{subequations}
\begin{align}
  & \mathcal{M}_{ g_{l,n,q} \to c_{l,m,q} }^{(i,j)} \left(c_{l,m,q} \right) \notag\\
  & \approx \!\!\int\!\! \mathcal{VM} \!\left( \!\Delta_{m,q} ; \eta_{ \Delta_{m,q} \to g_{l,n,q} }^{(i,j)} \!\right) 
  \mathcal{CN} \!\left(\! c_{l,m,q}  ; \mu_{l,m,q,n}^{(i,j)} , \big(\sigma^{(i,j)}_{l,m,q,n}\big)^{\!2}  \right) \notag\\
  & \;\;\;\;\; {\rm d} \Delta_{m,q}   \label{eq:message_I_g_c_int3}  \\
  & \approx \mathcal{CN}\left( c_{l,m,q} ; \mu_{ g_{l,n,q} \to c_{l,m,q} }^{(i,j)} , \big( \sigma^{(i,j)}_{ g_{l,n,q} \to c_{l,m,q} } \big)^2  \right) ,  \label{eq:message_I_g_c} 
\end{align}
\end{subequations}
where \eqref{eq:message_I_g_c} follows by applying the moment matching principle. 
The mean and variance are given by 
\begin{align}
  & \mu_{ g_{l,n,q} \to c_{l,m,q} }^{(i,j)}  
  = \bar{\mu}_{l,m,q,n}^{(i,j)} \notag\\
  & \;\;\;\;\; \times \exp \left(  \jmath (n - 1)  \mu_{  \Delta_{m,q} \to g_{l,n,q} }^{(i,j)}  \right) 
   \tfrac{I_{n-1} \left(   \kappa_{ \Delta_{m,q} \to g_{l,n,q} }^{(i,j)}   \right)}{I_0\left(  \kappa_{ \Delta_{m,q} \to g_{l,n,q} }^{(i,j)} \right)}  ,\label{eq:message_I_g_c_mu}
\end{align}
\begin{equation}
  \big(\sigma^{(i,j)}_{ g_{l,n,q} \to c_{l,m,q} } \big)^2 
  = \big(\sigma^{(i,j)}_{l,m,q,n}\big)^2 + |\bar{\mu}_{l,m,q,n}^{(i,j)}|^2 - |\mu_{ g_{l,n,q} \to c_{l,m,q} }^{(i,j)}|^2 . \label{eq:message_I_g_c_sigma}
\end{equation}



\subsubsection{Message from $c_{l,m,q}$ to $g_{l,n,q}$} 
The message from 
$c_{l,m,q}$ to 
$g_{l,n,q}$ is proportional to a Gaussian PDF as follows:
\begin{subequations} 
\begin{align}
  & \mathcal{M}_{  c_{l,m,q}  \to  g_{l,n,q} }^{(i,j+1)} \!\!\left( c_{l,m,q} \right) \notag\\
  & = \!\mathcal{M}_{   u_{l,m,q} \to c_{l,m,q}  }^{(i)} \!\!\left( c_{l,m,q} \right) 
  \!{\prod}_{n'\neq n} \!\!\! \mathcal{M}_{ g_{l,n,q} \to c_{l,m,q} }^{(i,j)} \!\!\left(c_{l,m,q} \right) \\
  & \propto \!\mathcal{CN}\!\left( c_{l,m,q} ; \mu_{  c_{l,m,q}  \to  g_{l,n,q} }^{(i,j+1)} , \big( \sigma^{(i,j+1)}_{  c_{l,m,q}  \to  g_{l,n,q} } \big)^2 \right) \!, \label{eq:message_I_c_g}
\end{align}  
\end{subequations}
{\blue where \eqref{eq:message_I_c_g} follows because $\mathcal{M}_{   u_{l,m,q} \to c_{l,m,q}  }^{(i)} \left( c_{l,m,q} \right) = \mathcal{CN} \big(  c_{l,m,q}  ; \mu_{   u_{l,m,q} \to c_{l,m,q}  }^{(i)} , \big( \sigma^{(i)}_{   u_{l,m,q} \to c_{l,m,q}  }\big)^2 \big) $ as will be introduced in \eqref{eq:message_I_u_c}.}  
The mean and variance in \eqref{eq:message_I_c_g} are given by
\begin{equation}
    \mu_{  c_{l,m,q}  \to  g_{l,n,q} }^{(i,j+1)}
    \!= \tfrac{ \frac{\mu_{   u_{l,m,q} \to c_{l,m,q}  }^{(i)}}{ \big( { \sigma^{(i)}_{   u_{l,m,q} \to c_{l,m,q}  } } \big)^{2} }     
    + \sum_{n'\neq n} \! \frac{ \mu_{   g_{l,n,q} \to c_{l,m,q}  }^{(i,j)} }{ \left({\sigma^{(i,j)}_{   g_{l,n,q} \to c_{l,m,q}  }} \right)^{2} } }{ \big( { \sigma^{(i)}_{   u_{l,m,q} \to c_{l,m,q}  } } \big)^{-2} \!+ \sum_{n'\neq n} \!  \left({\sigma^{(i,j)}_{   g_{l,n,q} \to c_{l,m,q}  }} \right)^{-2} } , \label{eq:message_I_c_g_mu}
\end{equation}
\begin{equation}
    \left( \!\sigma^{(i,j+1)}_{  c_{l,m,q}  \to  g_{l,n,q} } \!\right)^2 \!=\! \tfrac{ 1 }{ \left(\! {\sigma^{(i)}_{   u_{l,m,q} \to c_{l,m,q}  } }\! \right)^{\!-2} \! + \sum_{n'\neq n} \! \left(\! {\sigma^{(i,j)}_{   g_{l,n,q} \to c_{l,m,q}  }}\! \right)^{\!-2} } . \label{eq:message_I_c_g_sigma}
\end{equation}

\subsubsection{Message from $g_{l,n,q}$ to $\Delta_{m,q}$} The message from factor node $g_{l,n,q}$ to variable node $\Delta_{m,q}$ is expressed as
\begin{subequations}
\begin{align}
  & \mathcal{M}_{ g_{l,n,q} \to  \Delta_{m,q}  }^{(i,j+1)} \left( \Delta_{m,q} \right) \notag\\
  & \!= \!\int  \prod_{m'} \mathcal{M}_{  c_{l,m',q}  \to  g_{l,n,q} }^{(i,j+1)} \!\left( c_{l,m',q} \right)  
  \!\prod_{m'\neq m} \!\mathcal{M}_{ \Delta_{m',q} \to g_{l,n,q} }^{(i,j)} \!\left( \Delta_{m',q} \right) 
  \notag\\ 
  & \;\;\;\; \times p\!\left( \left. y_{l,n,q} \right| [\mathbf{C}_q]_{l,:}  ,  [\bm{\Delta}]_{:,q}  \right)  
  {\rm d} [\mathbf{C}_q]_{l,:}  {\rm d} [\bm{\Delta}]_{:,q \backslash \Delta_{m,q} }     \\
  & \!= \!\!\!\int \!\!\!\mathcal{M}_{  c_{l,m,q}  \to  g_{l,n,q} }^{(i,j+1)} \!\!\left( c_{l,m,q} \right) 
  \!f_n^{(i,j+1)} \!\!\left( c_{l,m,q} , \!\Delta_{m,q} \right) 
  \!{\rm d} c_{l,m,q} . \!\label{eq:message_I_g_delta_int}
\end{align}
\end{subequations}
Similar to \eqref{eq:message_I_g_c_int_part2_G}, we have
\begin{equation}
    f_n^{(i,j+1)}  \! \left( c_{l,m,q} , \!\Delta_{m,q} \right)
   \! \approx \!
    \mathcal{CN} \left(  c_{l,m,q}  ; \mu_{l,m,q,n}^{(i,j+1)} ,  \big( \sigma^{(i,j+1)}_{l,m,q,n} \big)^2  \right) .  \label{eq:message_I_g_delta_int_part2_G}
\end{equation}
with mean $\mu^{(i,j+1)}_{l,m,q,n} = \bar{\mu}_{l,m,q,n}^{(i,j+1)}  \exp\left( \jmath (n-1) \Delta_{m,q} \right)$. 
The parameter $\bar{\mu}_{l,m,q,n}^{(i,j+1)}$ and variance $\big(\sigma^{(i,j+1)}_{l,m,q,n}\big)^2$ are obtained by changing $\mu_{ c_{l,m,q} \to g_{l,n,q} }^{(i,j)}$ and $\big(\sigma^{(i,j)}_{ c_{l,m,q} \to g_{l,n,q} }\big)^2$ in \eqref{eq:message_I_g_c_int_part2_G_barmu} and \eqref{eq:message_I_g_c_int_part2_G_sigma2} to $\mu_{ c_{l,m,q} \to g_{l,n,q} }^{(i,j+1)}$ and $\big(\sigma^{(i,j+1)}_{ c_{l,m,q} \to g_{l,n,q} }\big)^2$, respectively. 
Then, we have
\begin{subequations}
\begin{align}
  & \mathcal{M}_{ g_{l,n,q} \to  \Delta_{m,q}  }^{(i,j+1)} \left( \Delta_{m,q} \right) \notag\\
  & = \mathcal{CN} \left( 0  ; \mu_{  c_{l,m,q}  \to  g_{l,n,q} }^{(i,j+1)} - {\bar{\mu}}_{l,m,q,n}^{(i,j+1)} \exp\left( \jmath (n-1) \Delta_{m,q} \right)   ,  \right. \notag\\
  & \;\;\;\;\;  \left. 
  \big( \sigma^{(i,j+1)}_{  c_{l,m,q}  \to  g_{l,n,q} } \big)^2  + \big( {\sigma}^{(i,j+1)}_{l,m,q,n}  \big)^2
    \right) \label{eq:message_I_g_delta_int_all_2} \\
  & \propto  
  \exp \left\{ \Re \left(  \left( \eta^{(i,j+1)}_{ g_{l,n,q} \to  \Delta_{m,q}  } \right)^{*} \exp\left( \jmath (n-1) \Delta_{m,q} \right)   \right) \right\} \label{eq:message_I_g_delta_int_all_3}\\ 
  & \propto \mathcal{VM}\left( (n-1)\Delta_{m,q} ; \eta^{(i,j+1)}_{ g_{l,n,q} \to  \Delta_{m,q}  } \right) , \label{eq:message_I_g_delta}
\end{align} 
\end{subequations}
following from Gaussian and VM PDFs with
\begin{equation}
    \eta_{ g_{l,n,q} \to  \Delta_{m,q}  }^{(i,j+1)} = \tfrac{ 2 \mu_{  c_{l,m,q}  \to  g_{l,n,q} }^{(i,j+1)} \left({\bar{\mu}}^{(i,j+1)}_{l,m,q,n} \right)^{*} } { \big( \sigma^{(i,j+1)}_{  c_{l,m,q}  \to  g_{l,n,q} } \big)^2 +  \big(\sigma^{(i,j+1)}_{l,m,q,n} \big)^2  } . \label{eq:message_I_g_delta_eta}
\end{equation}

\subsubsection{Message from $\Delta_{m,q}$ to $f_{m,q}$} The message from variable node $\Delta_{m,q}$ to factor node $f_{m,q}$ is expressed as 
\begin{subequations}
\begin{align}
  \mathcal{M}_{ \Delta_{m,q} \to f_{m,q} }^{(i,j+1)} \left( \Delta_{m,q} \right)   
  & = {\prod}_{l,n} \mathcal{M}_{ g_{l,n,q} \to \Delta_{m,q} }^{(i,j+1)} \left( \Delta_{m,q} \right) \label{eq:message_I_delta_fq_nl} \\
  & \approx \mathcal{VM} \left(  \Delta_{m,q}; \eta_{ \Delta_{m,q} \to f_{m,q} }^{(i,j+1)}  \right) . \label{eq:message_I_delta_fq}
\end{align} 
\end{subequations}
The term on the RHS of~\eqref{eq:message_I_delta_fq_nl} is similar to \eqref{eq:message_I_delta_g_VMnl} with the only difference of incorporating an additional message in the multiplication. 
Following similar approach as in \eqref{eq:message_I_delta_g_VMnl}-\eqref{eq:message_I_delta_g_VM}, we approximate \eqref{eq:message_I_delta_fq_nl} by a single VM PDF, as shown in \eqref{eq:message_I_delta_fq}.


\subsubsection{Message from $f_{m,q}$ to $\mathbf{p}_{{\rm t},m}$}
Following from the sift property of Dirac delta function under integral, the message from factor node $f_{m,q}$ to variable node $\mathbf{p}_{{\rm t},m}$ saisfies
\begin{subequations}
\begin{align}
  & \mathcal{M}_{ f_{m,q} \to  \mathbf{p}_{{\rm t},m} }^{(i,j+1)} \left( \mathbf{p}_{{\rm t},m} \right) \notag\\
  & = \int \mathcal{M}_{ \Delta_{m,q} \to f_{m,q} }^{(i,j+1)} \left( \Delta_{m,q} \right) 
  p\left( \left. \Delta_{m,q} \right| \mathbf{p}_{{\rm t},m} \right)
  {\rm d} \Delta_{m,q} \\
  & \propto \mathcal{VM} \!\left( \tfrac{\pi\left( y_{{\rm b},q} - y_{{\rm t},m} \right)}{ \left\| \mathbf{p}_{{\rm b},q} - \mathbf{p}_{{\rm t},m} 
 \right\|_2 } ; \eta_{ f_{m,q} \to  \mathbf{p}_{{\rm t},m} }^{(i,j+1)} \right) \!,  \label{eq:message_I_fq_pt} 
\end{align}
\end{subequations}
where $\eta_{ f_{m,q} \to  \mathbf{p}_{{\rm t},m} }^{(i,j+1)}$ is given by
\begin{equation} \label{eq:message_I_fq_pt_eta}
    \eta_{ f_{m,q} \to  \mathbf{p}_{{\rm t},m} }^{(i,j+1)} \!\!=\! \eta_{ \Delta_{m,q} \to f_{m,q} }^{(i,j+1)} \!\!=\!   \kappa_{ f_{m,q} \to  \mathbf{p}_{{\rm t},m} }^{(i,j+1)} \!\exp\!\left( \jmath \mu_{ f_{m,q} \to  \mathbf{p}_{{\rm t},m} }^{(i,j+1)} \right) . 
\end{equation} 

\textbf{ADCE Module}: {\blue
The ADCE module addresses the multi-user signal model $\mathbf{C}_q = \bar{\mathbf{X}} \mathbf{B}_q$, which is a linear combination of the transmitted pilot signals of active users. 
Based on the extrinsic messages over $\mathbf{C}_q$ from the TL module and the prior information on user activities $\alpha_k$'s and effective attenuation coefficients $s_{k,m,q}$'s, the ADCE module aims to update the estimates on $\alpha_k$'s and $s_{k,m,q}$'s, subsequently providing extrinsic messages over $\mathbf{C}_q$ for the TL module. 
Specifically, the ADCE module operates in three phases: the (into)-phase, the (within)-phase, and the (out)-phase. 
In the (into)-phase, the messages propagate through the paths $c_{l,m,q} \to u_{l,m,q}$ and $\alpha_k \to v_{k,m,q} \to b_{k,m,q}$ (see Lines 16-18 of Algorithm~\ref{alg:GAMP-MP}). 
The (within)-phase corresponds to refining the estimates of $c_{l,m,q}$ and $b_{k,m,q}$ via the GAMP algorithm (see Line 20 of Algorithm~\ref{alg:GAMP-MP}). 
In the (out)-phase, the updated messages of $b_{k,m,q}$ are used to update the beliefs about $\alpha_k$'s and $s_{k,m,q}$'s, and the estimate of $c_{l,m,q}$ is used to provide extrinsic messages for the TL module (see Lines 22-23 of Algorithm~\ref{alg:GAMP-MP}).} 
The messages in this module are described sequentially in the following.

\textit{1) Message from $c_{l,m,q}$ to $u_{l,m,q}$:} The message from variable node $c_{l,m,q}$ to factor node $u_{l,m,q}$ is proportional to a Gaussian density as follows: 
\begin{subequations}
\begin{align}
  & \mathcal{M}_{  c_{l,m,q} \to u_{l,m,q} }^{(i)} \!\! \left( c_{l,m,q} \right) \!=\! {\prod}_{n=1}^{N} \!\mathcal{M}_{ g_{l,q,n} \to c_{l,m,q} }^{(i, {\rm Iter}_{\rm max} )} \!\!\left( c_{l,m,q} \right)  \\
  & \propto  \mathcal{CN}\left( c_{l,m,q} ; \mu_{ c_{l,m,q}  \to u_{l,m,q} }^{(i)} , \big(\sigma^{(i)}_{ c_{l,m,q}  \to u_{l,m,q} } \big)^2  \right) , \label{eq:message_I_c_u}
\end{align}  
\end{subequations}
where $\mu_{ c_{l,m,q}  \to u_{l,m,q} }^{(i)}$ and $ \big( \sigma^{(i)}_{ c_{l,m,q}  \to u_{l,m,q} } \big)^2 $ are given by
\begin{equation}
    \mu_{ c_{l,m,q}  \to u_{l,m,q} }^{(i)}
    \!=\! \big( \sigma^{(i)}_{ c_{l,m,q}  \to u_{l,m,q} } \big)^{\!2} 
    \!\sum_{n=1}^{N} \! \frac{ \mu_{ g_{l,n,q} \to c_{l,m,q} }^{(i,{\rm Iter}_{\rm max})} }{ \big( \sigma^{(i,{\rm Iter}_{\rm max})}_{ g_{l,n,q} \to c_{l,m,q} } \big)^{\!2} } ,\label{eq:message_I_c_u_mu}
\end{equation}
\begin{equation}
    \big( \sigma^{(i)}_{ c_{l,m,q}  \to u_{l,m,q} } \big)^2 = \left( {\sum}_{n=1}^{N}  \big( \sigma^{(i,{\rm Iter}_{\rm max})}_{ g_{l,n,q} \to c_{l,m,q} } \big)^{-2}   \right)^{-1} . \label{eq:message_I_c_u_sigma}
\end{equation} 

\textit{2) Message from $\alpha_k$ to $v_{k,m,q}$:} The message from $\alpha_k$ to $v_{k,m,q}$ corresponds to a Bernoulli density given by
\begin{subequations}
\begin{align}
  & \mathcal{M}_{ \alpha_k \to v_{k,m,q} }^{(i)} \left( \alpha_k \right) \notag\\
  & \propto {\rm{Ber}}\!\left( \alpha_k ; \lambda \right) 
  {\prod}_{ (m',q') \neq (m,q) } {\rm{Ber}}\!\left(\! \alpha_k ; \lambda_{ v_{k,m',q'} \to \alpha_k }^{(i)}  \right)  \\
  & \propto {\rm{Ber}}\left( \alpha_k ; \lambda_{ \alpha_k \to v_{k,m,q} }^{(i)}  \right) ,  \label{eq:message_I_alpha_v}
\end{align}
\end{subequations}
where \eqref{eq:message_I_alpha_v} follows because all terms including both $\alpha_k$ and $1-\alpha_k$ equal to $0$ and $\lambda_{ \alpha_k \to v_{k,m,q} }^{(i)}$ in \eqref{eq:message_I_alpha_v} is given by
\begin{align}
    & \lambda_{ \alpha_k \to v_{k,m,q} }^{(i)} \notag\\
    & = \tfrac{\lambda   \prod\limits_{ (m',q') \neq (m,q)} \lambda_{ v_{k,m',q'} \to \alpha_k }^{(i)} }{ \lambda  \!\! \prod\limits_{ (m' ,q') \neq (m,q) }   \!\! \lambda_{ v_{k,m',q'} \to \alpha_k }^{(i)}
    + (1-\lambda)  \!\! \prod\limits_{ (m',q') \neq (m,q) }  \!\! ( 1-\lambda_{ v_{k,m',q'} \to \alpha_k }^{(i)}  ) }  . \label{eq:message_I_alpha_v_lambda}
\end{align}

\textit{3) Message from $v_{k,m,q}$ to $b_{k,m,q}$:}   
The message from factor node $v_{k,m,q}$ to variable node $b_{k,m,q}$ is characterized by a Bernoulli-Gaussian distribution as follows:
\begin{subequations}
\begin{align}
  & \mathcal{M}_{ v_{k,m,q} \to b_{k,m,q} }^{(i)} \left( b_{k,m,q} \right) \notag\\
  & = {\sum}_{ \alpha_k \in \{0,1\} } {\sum}_{ s_{k,m,q} \in \{0,1\} } 
  p\left( s_{k,m,q} \right)
  \mathcal{M}_{ \alpha_k \to v_{k,m,q} }^{(i)} \left( \alpha_k \right) \notag \\ 
  & \;\;\;\; \times  p\left( b_{k,m,q} \left| \alpha_k , s_{k,m,q} \right.  \right) \\
  & \propto (1- \xi_{ v_{k,m,q} \to b_{k,m,q} }^{(i)} ) \delta \left(   b_{k,m,q} \right)   \notag\\
  & \;\;\;\;
  + \xi_{ v_{k,m,q} \to b_{k,m,q} }^{(i)} \mathcal{CN}\left( b_{k,m,q} ; 0 , \sigma^2_{\rho,k,m,q} \right) ,\label{eq:message_I_v_b} 
\end{align} 
\end{subequations}
where the parameter $\xi_{ v_{k,m,q} \to b_{k,m,q} }^{(i)}$ is given by
\begin{equation}
    \xi_{ v_{k,m,q} \to b_{k,m,q} }^{(i)} = \lambda_{ \alpha_k \to v_{k,m,q} }^{(i)}  \xi . \label{eq:message_I_v_b_xi}
\end{equation}

\textit{4) Messages between $u_{l,m,q}$ and $b_{k,m,q}$:}
The message from $u_{l,m,q}$ to $b_{k,m,q}$ and that from $b_{k,m,q}$ to $u_{l,m,q}$ are given by 
\begin{align} 
  & \mathcal{M}_{ u_{l,m,q} \to b_{k,m,q} }^{(i)} \left( b_{k,m,q} \right) \notag\\
  & = \int {\prod}_{k' \neq k} \mathcal{M}_{ b_{k',m,q} \to u_{l,m,q} }^{(i)} \left( b_{k',m,q} \right)  
  \mathcal{M}_{  c_{l,m,q}  \to  u_{l,m,q} }^{(i)} \!\left( c_{l,m,q} \right)   \notag\\
  & \;\;\;\; \times  p\left( \left. c_{l,m,q} \right| [\mathbf{B}_q]_{:,m}   \right)   
  {\rm d} [\mathbf{B}_q]_{:,m \backslash b_{k,m,q} }  {\rm d}   c_{l,m,q}  ,  \label{eq:message_I_u_b}
\end{align}
\begin{align} 
  \mathcal{M}_{ b_{k,m,q} \to u_{l,m,q} }^{(i+1)} \left( b_{k,m,q} \right) 
  = & {\prod}_{l' \neq l} \mathcal{M}_{\! u_{l',m,q} \to  b_{k,m,q} }^{(i)} \!\left( b_{k,m,q} \right)   \notag\\
  & \times 
  \mathcal{M}_{ \! v_{k,m,q}  \to  b_{k,m,q} }^{(i)} \!\left( b_{k,m,q} \right),  \label{eq:message_I_b_u}
\end{align}
{\blue respectively. However, the calculation of above messages is intractable since $K$ and $L$ are large in massive communication. To this end, we resort to the GAMP framework \cite{ref:GAMP_propose} to simplify message passing, where the central limit theorem and second-order Taylor expansion are leveraged. 
Due to space limitations, we only introduce the outline of GAMP in the following. 
In the $j$-th iteration of GAMP, the estimate of the noiseless signal $c_{l,m,q}$ is updated. 
Based on the message from $b_{k,m,q}$'s, the distribution of $c_{l,m,q}$ is obtained as $\mathcal{CN} \left(  c_{l,m,q}  ; {\hat{p}}_{l,m,q}^{(i,j)} , \tau_{l,m,q}^{p,(i,j)}  \right)$ with mean and variance shown in~\cite[Eq. (2a)-(2b)]{ref:GAMP_propose}. 
Taking the prior information $\mathcal{M}_{  c_{l,m,q} \to u_{l,m,q} }^{(i)} \left( c_{l,m,q} \right)$ into consideration, the MMSE estimate ${\hat{c}}_{l,m,q}^{0,(i,j)}$ for $c_{l,m,q}$ and its variance $\tau_{l,m,q}^{c,(i,j)} $ are given in~\cite[Eq. (11)-(12)]{ref:Zhu1}. 
The mean and variance of each element in the residual signal are updated as~\cite[Line (12)-(13) in Alg.~1]{ref:Zhu1}. 
Then, the plug-in estimate $\hat{r}_{k,m,q}^{(i,j)}$ of $b_{k,m,q}$, modeled as $\hat{r}_{k,m,q}^{(i,j)} = b_{k,m,q} + n_{k,m,q}^{(i,j)}$ with $n_{k,m,q}^{(i,j)} \sim \mathcal{CN}\left( n_{k,m,q}^{(i,j)} ; 0, \tau_{k,m,q}^{r,(i,j)} \right)$, is updated as in~\cite[Eq. (4a)-(4b)]{ref:GAMP_propose}. 
To exploit the sparsity in both user activity and channel's angle-domain, the message $\mathcal{M}_{ v_{k,m,q} \to b_{k,m,q} }^{(i)} \left( b_{k,m,q} \right)$, which captures the Bernoulli distributions of $\alpha_k$'s and $s_{k,m,q}$'s, is incorporated in the GAMP algorithm as prior information. 
Then, we can obtain the explicit expression of the posterior mean ${\hat{b}}_{k,m,q}^{(i,j+1)}$ and variance $\tau_{k,m,q}^{b,(i,j+1)}$ of $b_{k,m,q}$, as shown in \cite[(14)-(19)]{ref:Zhu1}. 
The above GAMP estimation is calculated for several iterations to stabilize the iterative procedure of the whole algorithm. 
After convergence, the extrinsic messages of $c_{l,m,q}$ and $b_{k,m,q}$ are derived as $\mathcal{CN} \left(  c_{l,m,q}  ; {\hat{p}}_{l,m,q}^{(i)} , \tau_{l,m,q}^{p,(i)}  \right)$ and $\mathcal{CN}\left( b_{k,m,q} ; \hat{r}_{k,m,q}^{(i)}, \tau_{k,m,q}^{r,(i)} \right)$, respectively. 
} 

\textit{5) Message from $u_{l,m,q}$ to $c_{l,m,q}$:} 
The message from factor node $u_{l,m,q}$ to variable node $c_{l,m,q}$ can be obtained based on the estimation result of the GAMP algorithm as follows:
\begin{align}
  & \mathcal{M}_{   u_{l,m,q} \to c_{l,m,q}  }^{(i+1)} \left( c_{l,m,q} \right) \notag\\
  & = \mathcal{CN} \left(  c_{l,m,q}  ; 
  \mu_{   u_{l,m,q} \to c_{l,m,q}  }^{(i+1)} , \big( \sigma^{(i+1)}_{   u_{l,m,q} \to c_{l,m,q}  } \big)^2 \right) , \label{eq:message_I_u_c}
\end{align}
where $\mu_{   u_{l,m,q} \to c_{l,m,q}  }^{(i+1)} = {\hat{p}}_{l,m,q}^{(i)}$ and $\big( \sigma^{(i+1)}_{   u_{l,m,q} \to c_{l,m,q}  } \big)^2= \tau^{p,(i)}_{l,m,q}$. 

\textit{6) Message from $b_{k,m,q}$ to $v_{k,m,q}$:} 
The message from variable node $b_{k,m,q}$ to factor node $v_{k,m,q}$ is obtained as 
\begin{align}
   & \mathcal{M}_{ b_{k,m,q} \to v_{k,m,q} }^{(i+1)}  \left( b_{k,m,q} \right) \notag\\
   & = \mathcal{CN} \!\left( b_{k,m,q} ; \mu_{ b_{k,m,q} \to v_{k,m,q} }^{(i+1)}, \big(\sigma^{(i+1)}_{ b_{k,m,q} \to v_{k,m,q} }\big)^2 \right) ,  \label{eq:message_I_b_v}
\end{align}
where $\mu_{ b_{k,m,q} \to v_{k,m,q} }^{(i+1)} \!=\! {\hat{r}}_{k,m,q}^{(i)}$ and $\big(\sigma^{(i+1)}_{ b_{k,m,q} \to v_{k,m,q} }\big)^2 \!=\! \tau_{k,m,q}^{r,(i)}$. 

\textit{7) Message from $v_{k,m,q}$ to $\alpha_k$:} 
The message from factor node $v_{k,m,q}$ to variable node $\alpha_k$ is given by
\begin{equation}
  \mathcal{M}_{ v_{k,m,q} \to \alpha_k }^{(i+1)} \!\left( \alpha_k \right) 
  \propto {\rm{Ber}}\left( \alpha_k ; \lambda_{ v_{k,m,q} \to \alpha_k }^{(i+1)}  \right) , \label{eq:message_I_vkmq_alphak}
\end{equation}
where  
\begin{equation}  
  \lambda_{ v_{k,m,q} \to \alpha_k }^{(i+1)} 
  = \frac{ \left( 1 - \xi  \right)  a^{(i+1)}   +  \xi  b^{(i+1)}  }{  \left( 2 - \xi  \right)  a^{(i+1)}   +  \xi  b^{(i+1)}  } , \label{eq:message_I_vkmq_alphak_lambda}
\end{equation}
\begin{equation}
    a^{(i+1)} = \mathcal{CN} \left( 0 ; \mu_{ b_{k,m,q} \to v_{k,m,q} }^{(i+1)}, \big(\sigma^{(i+1)}_{ b_{k,m,q} \to v_{k,m,q} } \big)^2 \right),\label{eq:message_I_vkmq_alphak_lambda_a}
\end{equation}
\begin{equation}
    b^{(i+1)} \!\!= \! \mathcal{CN} \!\!\left( 0 ; \mu_{ b_{k,m,q} \to v_{k,m,q} }^{(i+1)}\!, \big(\sigma^{(i+1)}_{ \!b_{k,m,q} \to v_{k,m,q} }\big)^2 \!+\! \sigma^2_{\!\rho,k,m,q} \right)\!.\label{eq:message_I_vkmq_alphak_lambda_b}
\end{equation} 
{\blue The detailed proof of \eqref{eq:message_I_vkmq_alphak} is provided in Appendix~\ref{Appendix_proof_I_vkmq_alphak}.}

\subsection{Output} \label{Sec:output}

After the algorithm converges, the MMSE estimation for target locations, user activities, and channels can be approximated. 
In the following, we use the superscript ``$(\infty)$'' to indicate that the algorithm is converged. Combining the messages from $\left\{f_{m,q} : q\in[Q]\right\}$ and the prior $p\left( \mathbf{p}_{{\rm t},m} \right)$, we~have 
\begin{subequations}
\begin{align}
  & \mathcal{M}_{  \mathbf{p}_{{\rm t},m}  } \left( \mathbf{p}_{{\rm t},m} \right)  \notag\\
  & = p\left( \mathbf{p}_{{\rm t},m} \right) 
  {\prod}_{q=1}^{Q} \mathcal{M}_{ f_{m,q} \to  \mathbf{p}_{{\rm t},m} }^{(\infty)}  \left( \mathbf{p}_{{\rm t},m} \right) \\
  & \approx \mathcal{N} \left( \mathbf{p}_{{\rm t},m} ; \bm{\mu}_{{\rm t},m} , \mathbf{C}_{{\rm t},m} \right) 
    \mathcal{N} \left( \mathbf{p}_{{\rm t},m} ; {\tilde{\bm{\mu}}}_{{\rm t},m}^{(\infty)}  , {\tilde{\mathbf{C}}}_{{\rm t},m}^{(\infty)}   \right) \label{eq:message_I_xy_output_approG} \\
  & \propto \mathcal{N} \left( \mathbf{p}_{{\rm t},m} ; \hat{\bm{\mu}}_{{\rm t},m}^{(\infty)}  , \hat{\mathbf{C}}_{{\rm t},m}^{(\infty)}  \right) ,  \label{eq:message_I_xy_output}
\end{align}
\end{subequations}
where \eqref{eq:message_I_xy_output_approG} follows along similar lines as in \eqref{eq:message_I_xy_fq_approG}, 
and $\hat{\bm{\mu}}_{{\rm t},m}^{(\infty)}$ and $\hat{\mathbf{C}}_{{\rm t},m}^{(\infty)}$ in \eqref{eq:message_I_xy_output} are given by 
\begin{equation}
  \hat{\bm{\mu}}_{{\rm t},m}^{(\infty)} = \hat{\mathbf{C}}_{{\rm t},m}^{(\infty)} 
  \left( {\mathbf{C}}_{{\rm t},m}^{-1} \bm{\mu}_{{\rm t},m}  + \big({\tilde{\mathbf{C}}}_{{\rm t},m}^{(\infty)}\big)^{\!-1}  {\tilde{\bm{\mu}}}_{{\rm t},m}^{(\infty)} \right) \!,  \label{eq:message_I_xy_output_mu}
\end{equation} 
\begin{equation}
  \hat{\mathbf{C}}_{{\rm t},m}^{(\infty)}
  = \left(  {\mathbf{C}}_{{\rm t},m}^{-1} + \big({\tilde{\mathbf{C}}}_{{\rm t},m}^{(\infty)}\big)^{-1}  \right)^{-1}. \label{eq:message_I_xy_output_var}
\end{equation} 
Then, the estimate of the position of target $m$ is given by 
\begin{equation}
  \hat{\mathbf{p}}_{{\rm t},m} = \left[ \hat{x}_{{\rm t},m}  , \hat{y}_{{\rm t},m} \right]^T 
  = \hat{\bm{\mu}}_{{\rm t},m}^{(\infty)} .  \label{eq:message_I_xy_output_ptm}
\end{equation}  
The estimate $\hat{\theta}_{m,q}$ of the AoA $\theta_{m,q}$ can be obtained by substituting the estimated target position into \eqref{eq:sin_theta_target_BS}. 

For the activity detection, the product of all incoming messages at variable node $\alpha_k$ is given by 
\begin{equation}
  \mathcal{M}_{ \alpha_k } \!\!\left( \alpha_k \right)  
  \!=\! p\!\left( \alpha_k \right) \!\prod_{ m,q } \! \mathcal{M}_{ v_{k,m,q} \to \alpha_k }^{(\infty)} \!\!\left( \alpha_k \right) 
  \!\propto\! {\rm{Ber}}\!\left( \alpha_k ; \!\lambda_{ \alpha_k }  \right) \!, 
\end{equation} 
where the estimate of active probability is given by 
\begin{equation}
    \lambda_{ \alpha_k }  
    = \tfrac{\lambda \prod_{ m,q }  \lambda_{ v_{k,m,q} \to \alpha_k }^{(\infty)} }{  \lambda  \prod_{ m,q }   \lambda_{ v_{k,m,q} \to \alpha_k }^{(\infty)}
    +  (1-\lambda) \prod_{ m,q }  ( 1 - \lambda_{ v_{k,m,q} \to \alpha_k }^{(\infty)}  ) }  . \label{eq:output_alpha_lambda}
\end{equation}
If $\lambda_{ \alpha_k }$ is greater than a threshold $\lambda_{\rm{thre}}$, user $k$ is regarded as being active, i.e., $\forall k\in[K]$, 
\begin{equation}
  \hat{\alpha}_k = \left\{ 
             \begin{array}{lr}
             1 , &  \lambda_{ \alpha_k }  \geq \lambda_{\rm{thre}}, \\
             0 , &  \lambda_{ \alpha_k }  < \lambda_{\rm{thre}}. 
             \end{array} \right.   \label{eq:output_alpha}
\end{equation}

Denote the effective channel matrix as $\bar{\mathbf{H}} = \left[  \bar{\mathbf{H}}_1 , \ldots ,  \bar{\mathbf{H}}_Q \right]$. After the convergence of the proposed algorithm, the estimate of 
$b_{k,m,q}$ is denoted as $\hat{b}_{k,m,q}^{(\infty)}$. Then, the effective channel between user $k$ and BS $q$ is estimated as
\begin{equation}\label{eq:output_h}
    \hat{\mathbf{h}}_{k,q} = {\sum}_{m=1}^{M} \hat{b}_{k,m,q}^{(\infty)} \mathbf{v}^T\! \left( \hat{\theta}_{m,q} \right), ~\forall k, q . 
\end{equation}
Denote $\hat{\mathbf{H}}_q = \big[ \hat{\mathbf{h}}_{1,q} , \ldots , \hat{\mathbf{h}}_{K,q} \big]^T $ and $\hat{\mathbf{H}} = \big[ \hat{\mathbf{H}}_1 ,  \ldots , \hat{\mathbf{H}}_Q \big] $.

{\blue
\begin{Remark} \label{remark:cellfree}
  Compared to the collocated architecture, in which all antennas are co-located at a single BS, the advantages of the adopted cell-free architecture are as follows. 
  From the communication perspective, the activity detection performance is enhanced benefiting from the macro-diversity in the cell-free network~\cite{ref:CF1}. 
  From the sensing perspective, in the considered narrowband system, it is infeasible to localize targets with a single BS since the target's position cannot be determined using only an AoA. In contrast, the cell-free network enables target localization by extracting multiple AoAs from a target to all BSs. 
  Moreover, when the path between a target and a BS is obstructed, the collocated architecture fails to estimate the parameters of this target. 
  In contrast, in cell-free networks, even if the path to a BS is obstructed, this target can still be detected by other BSs exploiting the spatial diversity, thereby enhancing the target detection reliability~\cite{ref:CF2}. 
\end{Remark}
}


{\blue
\begin{Remark} \label{remark:dynamic}
  The proposed scheme can be extended to dynamic environments.  
  Specifically, in the scenario with relatively high mobility, the Doppler frequency shift can assist target localization \cite{ref:downlink-ZhangRY1,ref:downlink-ZhangRY2}. 
  To leverage this, we can incorporate Doppler shifts and velocities into the joint distribution \eqref{eq:joint_PDF}, based on which the factor graph can be modified. 
  Exploiting the duality between Doppler shifts and angles, we can derive message passing equations for this new model based on our framework.   
  Moreover, when target moves, we need to perform tracking. This can be achieved by recursively applying our scheme, where the position information from the previous frame provides prior information for the current one based on a probabilistic transition model characterizing target movement. 
\end{Remark}
  }


\section{Performance Analysis} \label{Sec:performance}

{\blue In this section, we first derive the BCRB to characterize the performance limit 
for the Bayesian target position estimator. 
Next, we analyze the SE in the asymptotic regime. 
The complexity analysis of the proposed algorithm is then provided.}

\subsection{Bayesian Cram{\'e}r-Rao bound} \label{Sec:BCRB}

{\blue
Let $\rho_{k,m,q}^{\rm R} = \Re\{\rho_{k,m,q}\}$ and $\rho_{k,m,q}^{\rm I} = \Im\{\rho_{k,m,q}\}$ denote the real and imaginary parts of the attenuation coefficient $\rho_{k,m,q}, \forall k,m,q$, respectively. 
By stacking $\rho_{k,m,q}^{\rm R}$'s, we form the vector $\bm{\rho}^{\rm R} \in \mathbb{R}^{KMQ}$ satisfying $[\bm{\rho}^{\rm R}]_{KM(q-1)+K(m-1)+k,1} = \rho_{k,m,q}^{\rm R}$; similarly, we obtain $\bm{\rho}^{\rm I} \in \mathbb{R}^{KMQ}$ for the imaginary parts.
To derive the BCRB, we consider the ideal case where $\alpha_k$'s and $s_{k,m,q}$'s are known in advance. 
Define $\bm{\psi} = [ \mathbf{p}_{{\rm t}}, ({\bm\rho}^{\rm R})^T, ({\bm\rho}^{\rm I})^T]^T$ as the collection of parameters to be estimated, where $\mathbf{p}_{{\rm t}} = [ \mathbf{p}_{{\rm t},1}^T, \ldots, \mathbf{p}_{{\rm t},M}^T]^T$. 
In the following, we characterize the BCRB 
based on the received signal $\mathbf{y} = \operatorname{vec}\left( \mathbf{Y} \right)
$ and prior distribution of $\bm{\psi}$.

To derive the BCRB, we first calculate the posterior Fisher information matrix (FIM) for estimating $\bm{\psi}$ as follows
\begin{equation}\label{eq:PFIM}
  \mathbf{F}_{\bm{\psi}} = \mathbf{F}_{\bm{\psi},0} + \mathbf{F}_{\bm{\psi},1}, 
\end{equation}
where $\mathbf{F}_{\bm{\psi},1}$ denotes the FIM corresponding to the prior knowledge about $\bm{\psi}$, 
given by
\begin{subequations}
\begin{align}
  \mathbf{F}_{\!\bm{\psi},1} \!\!
  & = \mathbb{E}_{\bm{\psi}} \!\Big[ \tfrac{\partial \log p(\bm{\psi}) }{\partial \bm{\psi}}   \Big( \tfrac{\partial \log p(\bm{\psi}) }{\partial \bm{\psi}} \Big)^T \Big] \label{eq:PFIM_pp0}\\
  & =  \!\operatorname{blkdiag}\!\left\{\;\!\! \mathbf{C}_{{\rm t}}^{-1}\! ,  2 (\operatorname{diag}\{ \;\!\!\bm{\sigma}_{\rho}^2 \;\!\!\})^{\!-1}\!, 2 (\operatorname{diag}\{\;\!\! \bm{\sigma}_{\rho}^2 \;\!\!\})^{\!-1} \;\!\!\right\}\!, \label{eq:PFIM_pp}
\end{align}
\end{subequations}
with $\mathbf{C}_{{\rm t}} = \operatorname{blkdiag}\left\{\mathbf{C}_{{\rm t},1}, \ldots, \mathbf{C}_{{\rm t},M} \right\}$ and $\bm{\sigma}_{\rho}^2 \in \mathbb{R}^{KMQ}$ obtained by stacking $\sigma_{\rho,k,m,q}^2$'s. 
The matrix $\mathbf{F}_{\bm{\psi},0}$ in \eqref{eq:PFIM} denotes the FIM corresponding to received signals given~by
\begin{subequations}
\begin{align}
  \mathbf{F}_{\bm{\psi},0} & = \mathbb{E}_{\mathbf{y},\bm{\psi}} \Big[ \tfrac{\partial \log p(\mathbf{y}|\bm{\psi}) }{\partial \bm{\psi}}  
  \left(\tfrac{\partial \log p(\mathbf{y}|\bm{\psi}) }{\partial \bm{\psi}}\right)^T \Big] \label{eq:PFIM_01} \\
  & = \frac{2}{ \sigma_z^2 }  \mathbb{E}_{\mathbf{p}_{\rm t}} \!\begin{bmatrix} 
        \bar{\mathbf{F}}_{11}  & \bm{0} & \bm{0} \\ 
        \bm{0} & \Re(\mathbf{F}_{22}) & -\!\Im(\mathbf{F}_{22}) \\
        \bm{0} & -\!\Im(\mathbf{F}_{22})^T & \Re(\mathbf{F}_{22}) 
      \end{bmatrix}, \label{eq:PFIM_0}
\end{align}
\end{subequations}
where
\begin{equation} 
  \bar{\mathbf{F}}_{11}  
  \!=\! \tfrac{\pi^2N(N-1)(2N-1)}{6} \sum_{q=1}^Q \!\mathbf{R}_q  \operatorname{diag}\{ \sigma^2_{c,m,q} , \ldots, \sigma^2_{c,M,q}  \}  \mathbf{R}_q^H  , \label{eq:PFIM_0_F11bar}
\end{equation}
\begin{equation} 
  \sigma^2_{c,m,q} = {\sum}_{l=1}^{L} {\sum}_{k=1}^{K}  \alpha_{k} \bar{x}_{k,l}^2
  s_{k,m,q} \sigma^2_{\rho,k,m,q} , \label{eq:PFIM_0_F11bar_sigmac}
\end{equation}
\begin{equation} 
  \mathbf{R}_q = \operatorname{blkdiag}\{ \mathbf{r}_{1,q}, \ldots, \mathbf{r}_{M,q} \}
   , \label{eq:PFIM_0_F11bar_Rq}
\end{equation}
\begin{equation} 
  \mathbf{r}_{m,q} = \left[ \tfrac{  \left( y_{{\rm b},q} - y_{{\rm t},m}\right) \left( x_{{\rm b},q} - x_{{\rm t},m} \right) }{ d_{m,q}^3 } , 
  \tfrac{  - \left( x_{{\rm b},q} - x_{{\rm t},m} \right)^2 }{d_{m,q}^3} \right]^T . \label{eq:PFIM_0_F11bar_rmq}
\end{equation} 
The expression of $\mathbf{F}_{ 22 } $ and the proof details for deriving \eqref{eq:PFIM_0} are shown in Appendix~\ref{Appendix_proof_CRB}.

Then, the BCRB for the MSE of any estimator of $\mathbf{p}_{\rm t}$, denoted by $\hat{\mathbf{p}}_{\rm t}$, can be expressed as  
\begin{equation}\label{eq:BCRB}
  \mathbb{E} \left[ \left\| \mathbf{p}_{\rm t} - \hat{\mathbf{p}}_{\rm t} \right\|_2^2 \right]
  \geq \operatorname{tr}\Big( \left(\tfrac{2}{ \sigma_z^2 } \mathbb{E}_{\mathbf{p}_{\rm t}} \! \left[\bar{\mathbf{F}}_{11} \right] + \mathbf{C}_{\rm t}^{-1}\right)^{-1} \Big) . 
\end{equation}    

}

\subsection{State Evolution} \label{Sec:SE}


{\blue 
When the elements in the pilot matrix satisfy i.i.d. sub-Gaussian distributions, the MSE of GAMP-based algorithms can be accurately tracked by a set of SE equations in the asymptotic regime with $L,K \to \infty$ and fixed $\frac{L}{K}$~\cite{ref:GAMP_propose}.

We consider the GAMP method with scalar variances for each target $m$ and BS $q$ as in \cite{ref:GAMP_propose}. These scalar variances at the $j$-th inner iteration of the $i$-th outer iteration are expressed as
\begin{equation}\label{eq:SE_tau_p}
  \tau_{m,q}^{p,(i,j)} = KP \tau_{m,q}^{b,(i,j)}, 
\end{equation}
\begin{equation}\label{eq:SE_tau_c}
  \tau_{m,q}^{c,(i,j)} = \frac{1}{L} \sum_{l=1}^{L} \tfrac{\tau_{m,q}^{p,(i,j)} \big( \sigma^{(i)}_{ c_{l,m,q}  \to u_{l,m,q} } \big)^2 }{\tau_{m,q}^{p,(i,j)} + \big(\sigma^{(i)}_{ c_{l,m,q}  \to u_{l,m,q} } \big)^2 }  , 
\end{equation}
\begin{equation}\label{eq:SE_tau_r}
  \tau_{m,q}^{r,(i,j)} = \tfrac{ (\tau_{m,q}^{p,(i,j)})^2 }{LP  \left( \tau_{m,q}^{p,(i,j)} - \tau_{m,q}^{c,(i,j)} \right) },  
\end{equation}
\begin{equation}\label{eq:SE_tau_b}
  \tau_{m,q}^{b,(i,j+1)} = \frac{1}{K} \sum_{k=1}^{K} \tau_{k,m,q}^{b,(i,j+1)} . 
\end{equation} 
Then, at each iteration, the asymptotic MSE for each target $m$ and BS $q$ can be tracked based on the following recursion~\cite{ref:GAMP_propose}
\begin{equation}
  \bar{\tau}_{m,q}^{b,(i,j)} \!\!=\!\! \frac{1}{K} \!\sum_{k=1}^{K} \! \mathbb{E} \!\left[ \operatorname{Var} \!\left( b_{k,m,q} | \hat{r}_{k,m,q}^{(i,j)}, \bar{\tau}_{m,q}^{r,(i,j)}, \xi_{ v_{k,m,q} \to b_{k,m,q} }^{(i)}   \right)  \right] \!,  \label{eq:SE_bar_tau_b}
\end{equation} 
\begin{equation}
  \bar{\tau}_{m,q}^{r,(i,j)} = \mathbb{E} \left[ \tfrac{ \left(\tau_{m,q}^{p,(i,j)}\right)^2 }{LP  \left( \tau_{m,q}^{p,(i,j)} - \tau_{m,q}^{c,(i,j)} \right) } \right], \label{eq:SE_bar_tau_r}
\end{equation} 
where the expectations are with respect to $\hat{r}_{k,m,q}^{(i,j)}$'s. 
As we can see, the probability $\xi_{ v_{k,m,q} \to b_{k,m,q} }^{(i)}$ related to the prior distribution of $b_{k,m,q}$ and the variance $\big( \sigma^{(i)}_{ c_{l,m,q}  \to u_{l,m,q} } \big)^2$ of the prior distribution of $c_{l,m,q}$ are incorporated in the SE. 
Since explicit expressions of these parameters are challenging to be obtained, we apply the Monte Carlo method to approximate~them.

Based on the SE, the asymptotic MSE for the estimation of $\mathbf{B}$ in each iteration can be tracked as $\mathbb{E} [ \| \hat{\mathbf{B}}^{(i,j)} - {\mathbf{B}}  \|_F^2  ] = \frac{1}{MQ}\sum_{m,q} \bar{\tau}_{m,q}^{b,(i,j)}$. This allows us to predict the MSE performance and validate the convergence behavior of the proposed scheme, as will be shown in Section \ref{Sec:simulation}. 

}

\subsection{Complexity Analysis}


The complexity of our algorithm is dominated by approximating the message $I_{  \mathbf{p}_{{\rm t},m} \to f_{m,q} }^{(i,j)} \left( \mathbf{p}_{{\rm t},m} \right), \forall m,q$ (in Line 5), approximating the message $I_{ \Delta_{m,q} \to g_{l,n,q} }^{(i,j)} \left( \Delta_{m,q} \right), \forall l,n,m,q$ (in Line 7), and running the GAMP algorithm (in Line 20). 
The complexity of the Gaussian approximation for $I_{  \mathbf{p}_{{\rm t},m} \to f_{m,q} }^{(i,j)} \left( \mathbf{p}_{{\rm t},m} \right), \forall m,q$ is $\mathcal{O}(n_1MQ^2)$, where $n_1$ denotes the iteration number of the Newton-Gauss method. 
According to the analysis in \cite{ref:Badiu}, the complexity of applying \cite[Algorithm~2]{ref:Badiu} to search for the most dominant component of the mixture of VM distributions is $\mathcal{O}(N^2)$. Thus, the update of $I_{ \Delta_{m,q} \to g_{l,n,q} }^{(i,j)} \left( \Delta_{m,q} \right), \forall l,n,m,q$ has complexity of $\mathcal{O}(N^3LMQ)$. 
For each $m$ and $q$, the complexity of the GAMP algorithm is $\mathcal{O}(n_2KL)$, where $n_2$ denotes its iteration number. 
In general, the complexity of the proposed algorithm per outer iteration is $\mathcal{O}( MQ (n_1Q+N^3L+n_2KL) )$.

\section{Numerical Results} \label{Sec:simulation}

In this section, we provide numerical examples to verify the effectiveness of the proposed algorithm. 
The system setup is as follows. 
The power spectral density of the AWGN at the BSs is $-169$~dBm/Hz and the channel bandwidth is $1$ MHz.  
There are $Q = 4$ BSs, which are located at $(0,50), (100,50), (50,0)$, and $(50,100)$ in meter, respectively, unless stated otherwise. 
There are $K=500$ users randomly localized in the network, and each of them becomes active with a probability of $\lambda = 0.2$. 
The position covariance matrix of target $m$ is set as $\mathbf{C}_{{\rm t},m} = \operatorname{diag}(1,1), \forall m$.
The variance of the attenuation coefficient of the path from user $k $ to target $m$ and then to BS $q$ 
is given by $\sigma^2_{\rho,k,m,q} = C_0 d_{k,m}^{-2} d_{m,q}^{-2}$, where $d_{k,m}$ and $d_{m,q}$ denote the distance between user $k$ and target $m$ and the distance between target $m$ and BS $q$, respectively, and $C_0 = -50$~dBm measures the path loss at the reference distance and the variance of RCS.


The estimation performance of target positions is evaluated in terms of the root MSE (RMSE) metric, defined as 
\begin{equation}
    {\rm RMSE} = \sqrt{ \mathbb{E} \left[ \tfrac{\sum_{m=1}^M \left\| \mathbf{p}_{{\rm t},m} - \hat{\mathbf{p}}_{{\rm t},m} \right\|_2^2}{M} \right] }. 
\end{equation}
The activity detection performance is evaluated by the probabilities of miss detection and false alarm, which are given~by
\begin{equation}
    P_{\rm MD} = \tfrac{| \mathcal{K}_a \backslash \hat{\mathcal{K}}_a |}{K_a}, \text{ and }
    P_{\rm FA} = \tfrac{| \hat{\mathcal{K}}_a \backslash \mathcal{K}_a |}{ |\hat{\mathcal{K}}_a| },
\end{equation}
respectively. Here, $\mathcal{K}_a$ and $\hat{\mathcal{K}}_a$ denote the true set and the estimated set of active users, respectively. 
We define the $K\times K$ diagonal matrix $\bm{\Lambda}_a$ satisfying $\left[\bm{\Lambda}_a\right]_{k,k} = 1$ if user $k$ is active and correctly detected, and $\left[\bm{\Lambda}_a\right]_{k,k} = 0$ otherwise. 
To evaluate the channel estimation performance, we adopt the normalized MSE~(NMSE) metric, which is defined as
\begin{equation}
    {\rm NMSE} = \mathbb{E} \left[ \tfrac{ \left\| \bm{\Lambda}_a \bar{\mathbf{H}}  - \bm{\Lambda}_a \hat{\mathbf{H}} \right\|_F^2 }{ \left\| \bm{\Lambda}_a \bar{\mathbf{H}}  \right\|_F^2 } \right].
\end{equation}

%
%
%

To show the effectiveness of the proposed method, we consider the following benchmark schemes:  
\begin{itemize}
    \item Benchmark Scheme I: 
    Under this scheme, we first perform ADCE based on the AMP algorithm with a soft-thresholding denoiser~\cite{ref:AMP,ref:Zhu-learning}. After the estimated channel $\hat{\mathbf{H}}$ is obtained, we apply the MUSIC algorithm to extract the AoA information, i.e. $\Delta_{m,q}$'s, from $\hat{\mathbf{H}}$. Last, we localize each target based on its AoAs to multiple BSs. Note that data association is a challenge for device-free networked sensing~\cite{ref:LLiu-ISAC,ref:QShi1,ref:association}. Here, we apply the data association method proposed in \cite{ref:association} to match the AoAs belonging to the same target for accurate~localization.

    \item Benchmark Scheme II: 
    We first apply the MUSIC algorithm to the received signals at each BS to estimate the AoA information, which is used to localize the targets applying the data association method \cite{ref:association}. Then, based on the estimated positions and the model in \eqref{eq:receive2}, the linear MMSE (LMMSE) principle is applied to obtain the estimate of $\mathbf{C}_q = \bar{\mathbf{X}} \mathbf{B}_q , \forall q$. 
    Then, we adopt the AMP algorithm with a soft-thresholding denoiser to estimate user activities $\alpha_k$'s and effective attenuation coefficients $\mathbf{B}_q$'s. 
    The effective channels are estimated applying \eqref{eq:output_h}.

    \item Benchmark Scheme III: 
    This scheme is similar to Scheme II, except that instead of using the AMP algorithm with a soft-thresholding denoiser, the GAMP-based ADCE module proposed in this work is adopted to exploit the statistical information of attenuation coefficients.

    \item Benchmark Scheme IV: 
    We first adopt the variational inference method in \cite{ref:Teng-Yuan,ref:Badiu} to exploit prior information of target positions for localization, where the VALSE algorithm is applied to approximate the message from $\Delta_{m,q}$ to $f_{m,q}$. 
    Then, as in Benchmark Scheme III, we adopt the GAMP-based ADCE module to output activity detection and channel estimation results. 

    \item {\blue Benchmark Scheme V: 
    We first discretize the angular domain into a set of grids as in \cite{ref:VBI}, thereby transforming AoA estimation at each BS to a compressed sensing problem. Following the block sparse Bayesian learning (SBL) method \cite{ref:bSBL}, the MAP estimate of the coefficient corresponding to each steering vector is obtained, with hyperparameters updated via expectation-maximum (EM). 
    Target positions are estimated after data association. 
    Then, based on the LMMSE estimate of $\mathbf{C}$, ADCE is performed following the block SBL method \cite{ref:bSBL}, where the variances of effective attenuation coefficients and LMMSE estimation errors are updated via~EM.     
    }
    
\end{itemize}
The computational complexities of benchmark schemes and the proposed scheme are summarized in Table~\ref{table:compare}. The complexity of our scheme is close to that of Benchmark Scheme IV. 
For Benchmark Schemes I, II, and III, the MUSIC algorithm employs a dense grid to discretize the continuous search space of possible AoAs, and the exhaustive search is required for data association, which results in high computational complexity, especially in the case with large $M$ and $Q$. 
{\blue For Benchmark Scheme V, the matrix inverse operation in EM-SBL and exhaustive search in data association lead to higher computational complexity than our scheme.}  

  \begin{table*}[!t] 
    \caption{Computational Complexity Comparison} \label{table:compare}
    \renewcommand{\arraystretch}{1}
    \centering
    \begin{threeparttable}          
    \begin{tabular}{c|c}
      \hline
      Algorithm & Complexity
      \\  \hline
      Benchmark Scheme I & $\mathcal{O}( n_2QLKN + N^2K_a + N^3 + GN^2(N-M) + n_1 (M!)^{Q-1} MQ  )$
      \\   
      Benchmark Schemes II and III & $\mathcal{O}( N^2L + N^3 + GN^2(N-M) + n_1 (M!)^{Q-1} MQ  + QLN^3 + n_2QLKM )$
      \\  
      Benchmark Scheme IV & $\mathcal{O}( n_1MQ^2 + n_4(QLNM^3 + QLN^2M) + n_1 (M!)^{Q-1} MQ  +  QLN^3 + n_2LKMQ )$
      \\ 
      Benchmark Scheme V & $\mathcal{O}( 
       (M')^2NQ + M'NLQ + n_5 QM'L   +  n_1 (M!)^{Q-1} MQ  + QLN^3 +  L^3 + K^2L + n_5 KLMQ   )$
      \\  
      Turbo-HyMP & $\mathcal{O}( n_3 ( n_1MQ^2 + QLMN^3 +n_2LKMQ ) ) $
      \\  \hline
    \end{tabular} 
    \end{threeparttable}     
    
    \vspace{1mm}
    {\scriptsize Note: $G$ denotes the number of grids in MUSIC, $n_1$ denotes the iteration number of Newton-Gauss, $n_2$ denotes the iteration number of GAMP/AMP, $n_3$ denotes the outer iteration number of the proposed algorithm, $n_4$ denotes the iteration number of VALSE, $n_5$ denotes the iteration number of EM, and $M'$ denotes the number of grids in SBL.} \vspace{-0.3cm}
  \end{table*}

\begin{figure}
    \centering
    \includegraphics[width=0.92\linewidth]{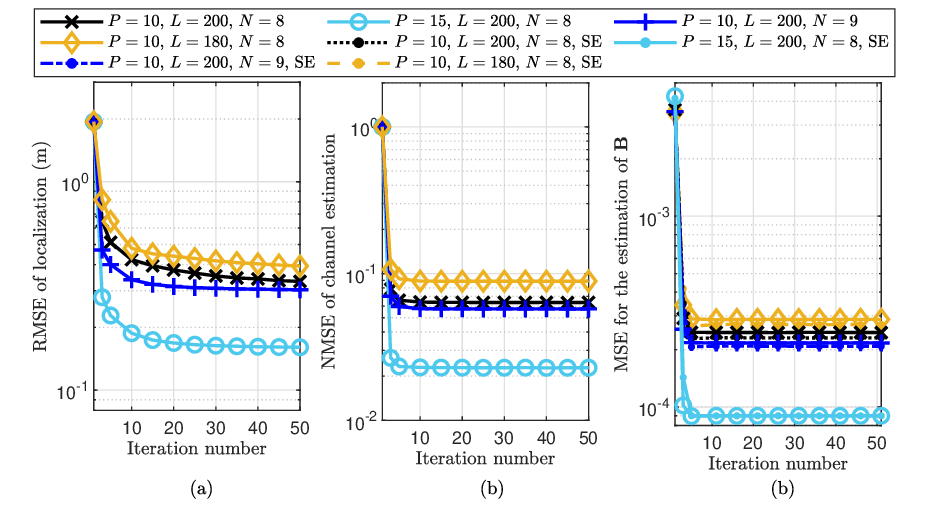}\\
    \vspace{-0.1cm}\caption{Convergence behavior versus the turbo iteration number with $
    M = 3$ and $\xi = 0.9$: (a)~Location RMSE of the proposed scheme; (b)~Channel estimation NMSE of the proposed scheme; and (c) MSE for effective channel attenuation coefficient estimation of the proposed scheme and its SE. 
} \label{fig:I} \vspace{-0.2cm}
\end{figure}

{\blue In Fig. \ref{fig:I}, we present the convergence behavior of the proposed algorithm in the scenario with $M = 3$ targets and path existence probability $\xi = 0.9$. 
It can be seen that the proposed Turbo-HyMP algorithm exhibits good convergence performance. Specifically, the target localization RMSE converges within about $20$ iterations, while the channel estimation NMSE converges significantly faster, requiring only about $5$ iterations. 
The estimation MSE of the effective channel attenuation matrix $\mathbf{B}$ can be accurately predicted by the SE, which also converges within $5$ iterations.  
Moreover, as the transmit power $P$, the number $N$ of receive antennas, or the pilot length $L$ increases, both the localization RMSE, channel estimation NMSE, and MSE for the estimation of $\mathbf{B}$ converge to lower values. Meanwhile, the convergence rate is slightly faster in the regime with larger $N$ and $P$.}

\begin{figure}
    \centering
    \includegraphics[width=0.92\linewidth]{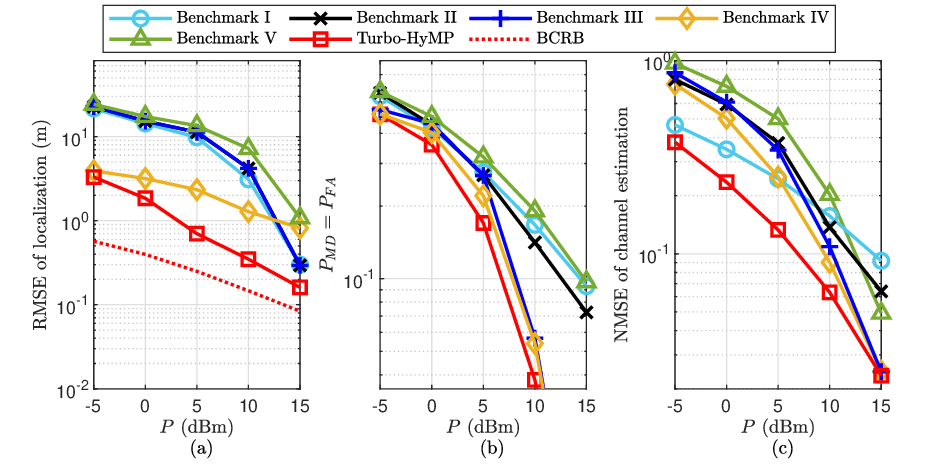}\\
    \vspace{-0.1cm}\caption{Target localization, activity detection, and channel estimation performance versus $P$ with $
    L = 200, M = 3, \xi = 0.9$, and $N = 8$: (a)~RMSE of location; (b)~$P_{\rm MD} = P_{\rm FA}$; and (c)~NMSE of channel estimation.
} \label{fig:P} \vspace{-0.2cm}
\end{figure}

Fig.~\ref{fig:P} shows the performance of the proposed algorithm and benchmark schemes versus transmit power $P$, 
assuming that the pilot length is $L=200$, the number of targets is $M = 3$, the number of receive antennas is $N = 8$, and the path existence probability is $\xi = 0.9$. 
For Benchmark Schemes I, II, and III, the number of searching grids in MUSIC is set to $G=10000$. 
It is shown that increasing $P$ significantly improves the joint estimation performance. 
We can observe that the TL accuracy of Benchmark Scheme II is close to that of Benchmark Scheme I, indicating that the sequence of performing TL or ADCE first has negligible impact on the localization performance of benchmarks in this case. 
For ADCE, Benchmark Scheme II has inferior performance to Benchmark Scheme I when $P$ is small, suffering from the accumulated AoA estimation error in the TL stage.  
However, in the regime with large $P$, Benchmark Scheme II outperforms Benchmark Scheme I since the AoA estimation accuracy is improved in this case. 
{\blue It is also shown that the proposed Turbo-HyMP algorithm significantly outperforms all benchmark schemes and achieves the localization RMSE much closer to the corresponding BCRB.} 
This is because Benchmark Schemes I-III adopt the MUSIC algorithm for localization, which can only provide limited AoA estimation performance when $P$ is insufficient. In contrast, the proposed algorithm can provide satisfactory performance even in the scenario with severe path loss attenuation. 
{\blue Benchmark Scheme V is inferior to our scheme due to inherent model mismatch induced by grid discretization and ineffective modeling of sparse priors using Gaussian distributions. 
Compared with all benchmark schemes, the proposed scheme has the advantage of fully leveraging the statistical dependency between the parameters, contributing to more accurate and reliable estimations.}
%


\begin{figure}
    \centering
    \includegraphics[width=0.92\linewidth]{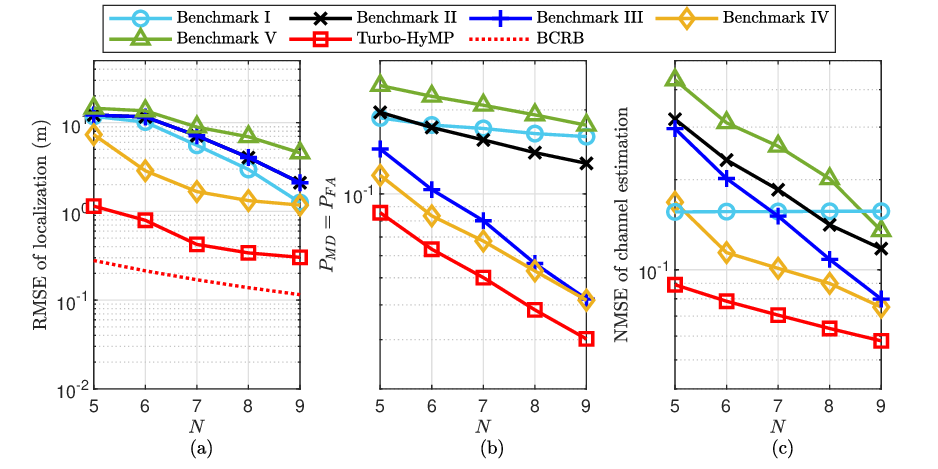}\\
    \vspace{-0.1cm}\caption{Target localization, activity detection, and channel estimation performance versus $N$ with $
    L = 200, M = 3, \xi = 0.9$, and $P = 10$~dBm: (a)~RMSE of location; (b)~$P_{\rm MD} = P_{\rm FA}$; and (c)~NMSE of channel estimation. 
} \label{fig:N} \vspace{-0.1cm}
\end{figure}

Fig.~\ref{fig:N} shows the 
performance of the proposed algorithm and benchmark schemes versus the number $N$ of receive antennas, 
in the scenario with $P = 10$~dBm, $M = 3$, $L=200$, and $\xi = 0.9$. 
Fig.~\ref{fig:N}(a) shows that the localization performance of Benchmark I is superior to that of Benchmark II in the considered regime, indicating that performing ADCE first is beneficial for the subsequent TL. 
Fig.~\ref{fig:N}(b) and Fig.~\ref{fig:N}(c) show that when $N$ is small, Benchmark I outperforms Benchmark II in terms of ADCE, considering that the ADCE performance of Benchmark II is limited by inaccurate AoA estimation in the TL stage. In contrast, when $N$ becomes large, the localization performance of Benchmark II is significantly improved, which further enhances the subsequent ADCE performance. These results indicate that to achieve better ADCE performance, directly performing ADCE is preferable for small $N$, whereas prioritizing TL over ADCE is more advantageous when $N$ is large. 
It is also shown that Benchmark III significantly outperforms Benchmark II in terms of ADCE by exploiting statistical information of attenuation coefficients in the second stage. 
Benchmark IV further incorporates prior knowledge of target positions, enabling it to achieve better TL and ADCE performance than Benchmark III. 
{\blue Benchmark V performs poorly due to model mismatch, ineffective sparse priors, and insufficient leverage of statistical dependency between unknown parameters.} 
Notably, the proposed Turbo-HyMP algorithm demonstrates a substantial performance improvement over all benchmark schemes. 
In particular, it exhibits significant superiority in target localization, with an overall $6$~dB enhancement compared with sequential counterparts. 
This performance gain can be interpreted as the ability of our scheme to efficiently exploit 
the statistical dependency between the unknown to be estimated, compared to benchmarks.


\begin{figure}
    \centering
    \includegraphics[width=0.92\linewidth]{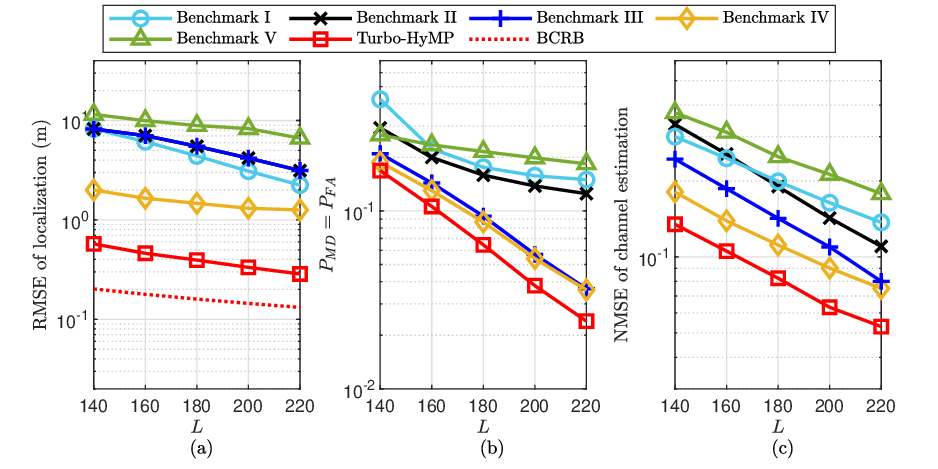}\\
    \vspace{-0.1cm}\caption{Target localization, activity detection, and channel estimation performance versus $L$ with $
    N = 8, M = 3, \xi = 0.9$, and $P = 10$~dBm: (a)~RMSE of location; (b)~$P_{\rm MD} = P_{\rm FA}$; and (c)~NMSE of channel estimation.} \label{fig:L} \vspace{-0.2cm}
\end{figure}

In Fig.~\ref{fig:L}, we compare the joint estimation performance achieved by the proposed algorithm and benchmark schemes under different pilot length $L$. 
Here, we assume $M = 3$, $N=8$, $\xi=0.9$, and $P = 10$~dBm. 
It is shown that compared to benchmarks, the proposed algorithm consistently achieves about $6$~dB reduction in localization error, as well as $1.5$~dB improvement in ADCE performance. 
Moreover, for ADCE, the performance of all schemes is significantly improved by increasing $L$ due to the enlarged number of measurements, which is similar to the results in conventional massive communication works \cite{ref:Liu-massive,ref:Zhu-learning,ref:Ke-EM}. 
Meanwhile, the TL performance also benefits from larger $L$ values. 
The above results validate the feasibility of leveraging rich pilot resources in massive communication to enhance the TL performance.


\begin{figure}
    \centering
    \includegraphics[width=0.92\linewidth]{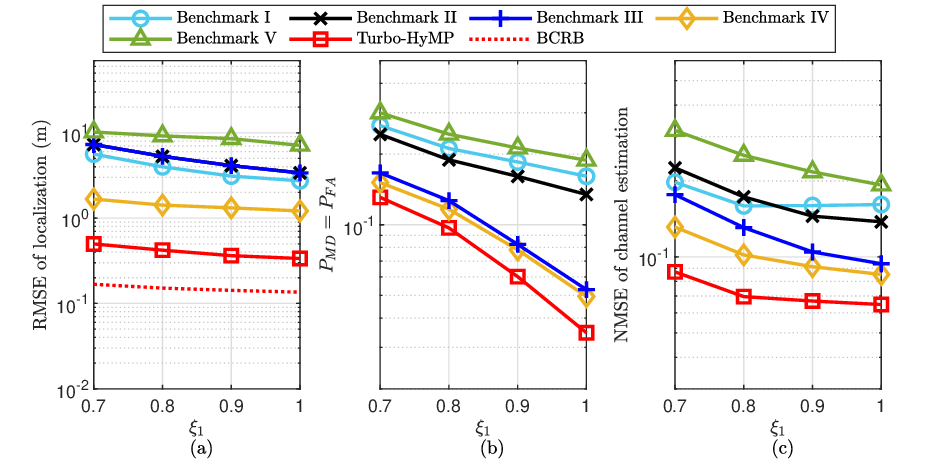}\\
    \vspace{-0.1cm}\caption{Target localization, activity detection, and channel estimation performance versus user-target path existing probability $\xi_1$ with $
    N = 8, M = 3, L = 200, \xi_2 = 0.92$, and $P = 10$~dBm: (a)~RMSE of location; (b)~$P_{\rm MD} = P_{\rm FA}$; and (c)~NMSE of channel estimation.} \label{fig:xi} \vspace{-0.2cm}
\end{figure}

In Fig.~\ref{fig:xi}, we compare the TL and ADCE performance of the proposed scheme and benchmarks versus user-target path existing probability $\xi_1$ while keeping the target-BS path existing probability $\xi_2$ fixed, such that the user-target-BS connectivity probability is expressed as $\xi = \xi_1\xi_2$. 
We assume $M = 3$, $N=8$, $L=200$, and $P = 10$~dBm. 
Fig.~\ref{fig:xi}(b) shows that as $\xi_1$ increases, each user obtains a higher number of successful access paths to the BS, thereby contributing to a significant reduction in activity detection error probability. 
In contrast, the performance gains in TL and channel estimation, shown in Fig.~\ref{fig:xi}(a) and Fig.~\ref{fig:xi}(c) respectively, are more moderate compared to activity detection. 
This occurs because while the increased number of effective paths enhances the received signal-to-noise ratio (SNR), it simultaneously introduces more fading coefficients to be estimated. These two competing effects result in less pronounced performance improvements for TL and channel estimation.

\begin{figure}
    \centering
    \includegraphics[width=0.92\linewidth]{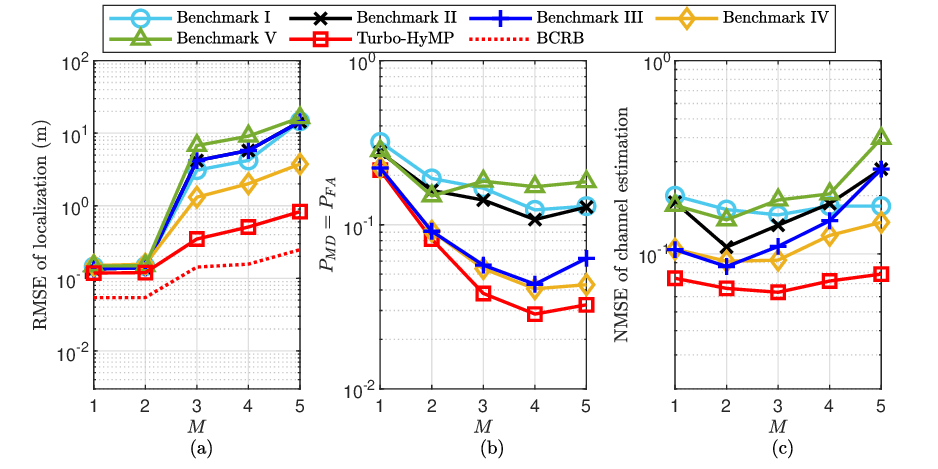}\\
    \vspace{-0.1cm}\caption{Target localization, activity detection, and channel estimation performance versus $M$ with $
    N = 8, L = 200, \xi = 0.9$, and $P = 10$~dBm: (a)~RMSE of location; (b)~$P_{\rm MD} = P_{\rm FA}$; and (c)~NMSE of channel estimation.} \label{fig:M} \vspace{-0.2cm}
\end{figure}

Fig.~\ref{fig:M} shows the TL and ADCE performance achieved by the proposed algorithm and benchmarks versus the number $M$ of targets, assuming $L=200$, $N = 8$, $P = 10$~dBm, and $\xi = 0.9$. 
It can be observed from Fig.~\ref{fig:M}(a) that the TL performance deteriorates as $M$ increases, suffering from the additional introduced AoAs to be estimated. 
Fig.~\ref{fig:M}(b) and Fig.~\ref{fig:M}(c) show that the ADCE performance initially improves but then deteriorates as $M$ increases. 
This is because for small $M$, adding more targets can provide more effective paths between users and BSs and enhance the received SNR. However, when $M$ is large, the AoA estimation error becomes the bottleneck, leading to performance degradation with increasing $M$. 
Our proposed algorithm demonstrates consistent superiority over benchmark schemes, especially in the case with large $M$. 
Our simulation results verify that the proposed scheme serves as an effective solution to enhance both the massive communication and TL functionalities in the considered ISAC system.

{\blue 
Fig.~\ref{fig:BSdeployment} presents the TL and ADCE performance achieved by the proposed scheme versus transmit power $P$ under two different BS deployment configurations: BSs are located at $(0,50), (100,50), (50,0)$, and $(50,100)$ in the first BS deployment configuration and located at $(0,70), (100,30), (30,0)$, and $(70,100)$ in the second one. 
We assume that $M = 3$, $L=200$, $N = 8$, and $\xi = 1$. 
It can be observed that the first deployment yields better performance than the second one. 
This is because target positions are estimated based on angular resolution, whose performance degrades as the distance between the BS and target increases. 
Compared to the second BS deployment configuration, the BSs under the first configuration provide superior resolution within the square region of interest, leading to enhanced target localization accuracy and consequently better ADCE performance. 
The performance comparison under different configurations provides guidance for BS deployment in practical systems. 

\begin{figure}
    \centering
    \includegraphics[width=0.92\linewidth]{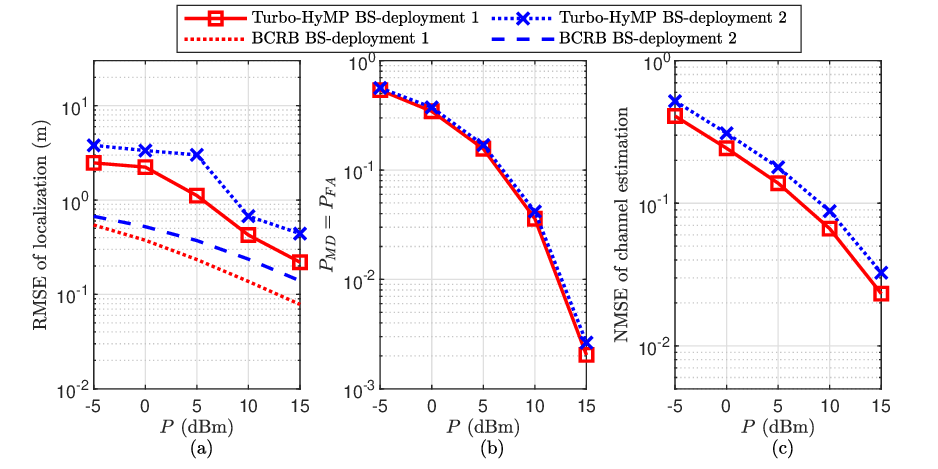}\\
    \vspace{-0.1cm}\caption{Target localization, activity detection, and channel estimation performance versus $P$ in two BS deployment configurations with $L = 200, M = 3, \xi = 1$, and $N = 8$: 
    (a)~RMSE of location; (b)~$P_{\rm MD} = P_{\rm FA}$; and (c)~NMSE of channel estimation.
} \label{fig:BSdeployment} \vspace{-0.2cm}
\end{figure}           
}

\section{Conclusion} \label{ref:conclusion}

  This paper considered the integrated massive communication and target localization problem in cell-free networks, where a random subset of users becomes active and their transmitted pilot signals are scattered by the targets and then received at the BSs. 
  In the considered problem, the user activities, effective channels, and target positions are coupled together, forming a complicated problem which cannot be jointly solved applying existing schemes directly. 
  We proposed a novel hybrid message passing-based framework to tackle the above challenge. 
  Specifically, we built a probabilistic model to characterize the intricate dependency between these parameters. 
  Then, we developed a Turbo-HyMP algorithm to jointly perform TL and ADCE, 
  where various techniques were applied to approximate the messages in the factor graph with efficient computations. 
  Theoretical analysis and numerical results were provided to verify the effectiveness of the proposed scheme in terms of activity detection, channel estimation, and target localization performance, indicating the feasibility of integrating massive communication and target localization in wireless networks with centralized joint processing of multiple remote radio units, which is an important candidate architecture for 6G systems. 
  {\blue In the future, it is interesting to 
  consider other practical issues in our approach, e.g., moving targets, user and BS synchronization issue, etc. 
  }


 \appendices

 \section{Derivation of Mean and Variance in \eqref{eq:message_I_xy_fq_approG}} \label{Appendix_proof_I_xy_fq}

 {\blue In the following, we omit the iteration index for brevity. 
 Denote $d_{m,q'} = \left\| \mathbf{p}_{{\rm b},q'} \!-\! \mathbf{p}_{{\rm t},m} \right\|_2 $ and $g_{q'}\left( \mathbf{p}_{{\rm t},m} \right) = \tfrac{\pi\left( y_{{\rm b},q'} - y_{{\rm t},m} \right)}{ d_{m,q'} }$. 
 The second term on the RHS of~\eqref{eq:message_I_xy_fq_prod} satisfies
 \begin{subequations}
 \begin{align}
   & {\prod}_{q'\neq q} \mathcal{M}_{ f_{m,q'} \to  \mathbf{p}_{{\rm t},m} } \left( \mathbf{p}_{{\rm t},m} \right) \notag\\
   & \propto e^{    \sum_{q'\neq q}   \kappa_{f_{m,q'} \to  \mathbf{p}_{{\rm t},m} }  
    \cos \left( g_{q'}\left( \mathbf{p}_{{\rm t},m} \right)  - \mu_{f_{m,q'} \to  \mathbf{p}_{{\rm t},m} } \right)  }\label{eq:proof_message_I_xy_fq_NG}  \\
   & \approx e^{ \sum_{q'\neq q} 
   \kappa_{f_{ m,q'} \to  \mathbf{p}_{{\rm t},m} }
   \left( 1- \frac{1}{2} \left(  g_{q'} \left( \mathbf{p}_{{\rm t},m} \right)  - \mu_{f_{m,q'} \to  \mathbf{p}_{{\rm t},m} } \right)^{ 2} \right)  } \label{eq:proof_message_I_xy_fq_NG2} \\ 
   & \propto e^{ -\frac{1}{2}  \left(  \mathbf{g}_q \left( \mathbf{p}_{{\rm t},m} \right) - {\bm{\mu}}_{m,q} \right)^T \mathbf{B}_{m,q} \left(  \mathbf{g}_q \left( \mathbf{p}_{{\rm t},m} \right) - {\bm{\mu}}_{m,q} \right) } \label{eq:proof_message_I_xy_fq_NG3} ,
 \end{align} 
 \end{subequations}
 where \eqref{eq:proof_message_I_xy_fq_NG} follows from~\eqref{eq:VMpdf1} and \eqref{eq:message_I_fq_pt}; 
 \eqref{eq:proof_message_I_xy_fq_NG2} follows by applying the Taylor approximation $\cos x \approx 1 - \frac{1}{2}x^2$; 
 and $\mathbf{g}_q\left( \mathbf{p}_{{\rm t},m} \right)$, $\mathbf{B}_{m,q}$, and ${\bm{\mu}}_{m,q}$ in \eqref{eq:proof_message_I_xy_fq_NG3} are given by 
 \begin{align}
   \mathbf{g}_q \left( \mathbf{p}_{{\rm t},m} \right) = & \left[ g_{1}\!\left( \mathbf{p}_{{\rm t},m} \right) , \ldots ,  g_{q-1}\!\left( \mathbf{p}_{{\rm t},m} \right),  \right. \notag\\
   & \;   \left. g_{q+1}\!\left( \mathbf{p}_{{\rm t},m} \right), \ldots, g_{Q}\!\left( \mathbf{p}_{{\rm t},m} \right) \right]^T \in \mathbb{R}^{Q-1}, \label{eq:proof_message_I_xy_fq_gq} 
 \end{align}
 \begin{align}
   \mathbf{B}_{m,q} = \operatorname{diag} & \left\{ \kappa_{f_{m,1} \to  \mathbf{p}_{{\rm t},m} }, \ldots, \kappa_{f_{m,q-1} \to  \mathbf{p}_{{\rm t},m} } ,  \right. \notag\\
   & \;  \left. \kappa_{f_{m,q+1} \to  \mathbf{p}_{{\rm t},m} } , \ldots, \kappa_{f_{m,Q} \to  \mathbf{p}_{{\rm t},m} } \right\}, \label{eq:proof_message_I_xy_fq_Bq} 
 \end{align}
 \begin{align}
   {\bm{\mu}}_{m,q} = & \left[ \mu_{f_{m,1} \to  \mathbf{p}_{{\rm t},m} } , \ldots, 
   \mu_{f_{m,q-1} \to  \mathbf{p}_{{\rm t},m} } ,  \right. \notag\\
   & \; \left. \mu_{f_{m,q+1} \to  \mathbf{p}_{{\rm t},m} } , \ldots, \mu_{f_{m,Q} \to  \mathbf{p}_{{\rm t},m} } \right]^T \in \mathbb{R}^{Q-1}. \label{eq:proof_message_I_xy_fq_muq} 
 \end{align}

 Denote the negative of the exponential term on the RHS of \eqref{eq:proof_message_I_xy_fq_NG3} as $f_q\left( \mathbf{p}_{{\rm t},m} \right)$. 
 We adopt the Newton-Gauss method to find the local minimum of $f_q\left( \mathbf{p}_{{\rm t},m} \right)$. 
 We iteratively calculate~\cite{ref:Newton_Gauss} 
 \begin{align}
     & \hat{\mathbf{p}}_{{\rm t},m,(i+1)} \notag\\
     & = \hat{\mathbf{p}}_{{\rm t},m,(i)}  
     - \left(  \mathbf{J}^T_{m,q} |_{ \mathbf{p}_{{\rm t},m} = \hat{\mathbf{p}}_{{\rm t},m,(i)} } \mathbf{B}_{m,q} \mathbf{J}_{m,q}|_{ \mathbf{p}_{{\rm t},m} = \hat{\mathbf{p}}_{{\rm t},m,(i)} } \right)^{-1}  \notag\\
     & \;\;\; \times \mathbf{J}^T_{m,q} |_{ \mathbf{p}_{{\rm t},m} = \hat{\mathbf{p}}_{{\rm t},m,(i)} } \mathbf{B}_{m,q} 
     \left(  \mathbf{g}_q\left( \hat{\mathbf{p}}_{{\rm t},m,(i)} \right) - {\bm{\mu}}_{m,q} \right) , \label{eq:proof_I_xy_fq_mut}
 \end{align}
 where $i$ denotes the iteration index and the Jacobi matrix $\mathbf{J}_{m,q}$ is given by
 \begin{equation}
     \mathbf{J}_{m,q} =  
     \left[ \tfrac{\partial \mathbf{g}_{q}\left( \mathbf{p}_{{\rm t},m} \right)  }{ \partial x_{{\rm t},m}}  \;\;\;\;\;
     \tfrac{\partial \mathbf{g}_{q}\left( \mathbf{p}_{{\rm t},m} \right)  }{ \partial y_{{\rm t},m}}  \right],  \label{eq:proof_message_I_xy_fq_Jmq} 
 \end{equation}  
 \begin{equation}
   \tfrac{\partial g_{q'}\left( \mathbf{p}_{{\rm t},m} \right)  }{ \partial x_{{\rm t},m}}  
   = \tfrac{  \pi\left( y_{{\rm b},q'} - y_{{\rm t},m}\right) \left( x_{{\rm b},q'} - x_{{\rm t},m} \right) }{ d_{m,q'}^3 }, \label{eq:proof_message_I_xy_fq_gq_partialx} 
 \end{equation}
 \begin{equation}
   \tfrac{\partial g_{q'}\left( \mathbf{p}_{{\rm t},m} \right)  }{ \partial y_{{\rm t},m}}  
   = \tfrac{  -\pi \left( x_{{\rm b},q'} - x_{{\rm t},m} \right)^2 }{d_{m,q'}^3}. \label{eq:proof_message_I_xy_fq_gq_partialy} 
 \end{equation}

 After obtaining the local minimizer $\hat{\mathbf{p}}_{{\rm t},m}$, we perform a second-order Taylor approximation for $f_q\left( \mathbf{p}_{{\rm t},m} \right)$ around $\hat{\mathbf{p}}_{{\rm t},m}$  
 \begin{align}
     &  f_q\!\left( \mathbf{p}_{{\rm t},m} \right) 
     \approx \!f_q\!\left( \hat{\mathbf{p}}_{{\rm t},m} \right)  
     \!+\! \Big[ \left. \nabla \! f_q\!\left( \hat{\mathbf{p}}_{{\rm t},m} \right)  \right|_{\mathbf{p}_{{\rm t},m} = \hat{\mathbf{p}}_{{\rm t},m}}  \Big]^T  
     \!\!\!\left( \mathbf{p}_{{\rm t},m} \!\!-\! \hat{\mathbf{p}}_{{\rm t},m} \right) \notag\\
     &  \;\;\;
     + \frac{1}{2}\left( \mathbf{p}_{{\rm t},m} \!-\! \hat{\mathbf{p}}_{{\rm t},m} \right)^T 
     \!  \left.  \mathbf{H}_{m,q} \right|_{\mathbf{p}_{{\rm t},m} = \hat{\mathbf{p}}_{{\rm t},m}} \!
     \left( \mathbf{p}_{{\rm t},m} \!-\! \hat{\mathbf{p}}_{{\rm t},m} \right) , \label{eq:proof_message_I_xy_fq_fq_Taylor} 
 \end{align}
 where $\nabla  f_q\left( {\mathbf{p}}_{{\rm t},m} \right) \approx \mathbf{0}$ in the case of $\mathbf{p}_{{\rm t},m} = \hat{\mathbf{p}}_{{\rm t},m}$ and the Hessian matrix satisfies $\mathbf{H}_{m,q} \approx  \mathbf{J}^T_m \mathbf{B}_m \mathbf{J}_m$. 
 Thus, we approximate the RHS of~\eqref{eq:proof_message_I_xy_fq_NG3} as a Gaussian PDF with mean ${\tilde{\bm{\mu}}}_{{\rm t},m,q} = \hat{\mathbf{p}}_{{\rm t},m}$ and variance ${\tilde{\mathbf{C}}}_{{\rm t},m,q} = \left( \mathbf{J}^T_{m,q} |_{ \mathbf{p}_{{\rm t},m} = \hat{\mathbf{p}}_{{\rm t},m } } \mathbf{B}_{m,q} \mathbf{J}_{m,q} |_{ \mathbf{p}_{{\rm t},m} = \hat{\mathbf{p}}_{{\rm t},m } } \right)^{-1}$. 
 }

 \section{Proof of \eqref{eq:message_I_vkmq_alphak}} \label{Appendix_proof_I_vkmq_alphak}

 {\blue In this appendix, we omit the iteration index for brevity. 
 Applying the sum-product rules, we have
 \begin{subequations}
 \begin{align}
   & \mathcal{M}_{ v_{k,m,q} \to \alpha_k } \left( \alpha_k \right) \notag\\
   &=\!\int {\sum}_{s_{k,m,q}\in\{0,1\}} \! \mathcal{M}_{ b_{k,m,q} \to v_{k,m,q} } \left( b_{k,m,q} \right)  
   p \left( s_{k,m,q} \right) \notag\\
   & \;\;\;\;\; \times p\left( b_{k,m,q} \left| \alpha_k , s_{k,m,q} \right.  \right)  {\rm d} b_{k,m,q} \label{eq:proof_message_I_vkmq_alphak_1} \\ 
   & = \!\int\!  {\sum}_{\!s_{k,m,q}\in\{0,1\}} \!\! \mathcal{CN} \! \left(\! b_{k,m,q} ; \mu_{ b_{k,m,q} \to v_{k,m,q} }\!, \sigma^2_{ b_{k,m,q} \to v_{k,m,q} } \right) \notag\\
   & \;\;\;\;\; \times 
   {\rm{Ber}} \left( s_{k,m,q} ; \xi  \right) 
   \left( (1-\alpha_k s_{k,m,q}) \delta \left(   b_{k,m,q} \right)   \right. \notag\\
   & \;\;\;\;\;  
   \left. + \alpha_k  s_{k,m,q} \;\! \mathcal{CN}\left( b_{k,m,q} ; 0 , \sigma^2_{\rho,k,m,q} \right)   \right) 
   {\rm d} b_{k,m,q}.  \label{eq:proof_message_I_vkmq_alphak}
 \end{align}
 \end{subequations}

 The term in the RHS of \eqref{eq:proof_message_I_vkmq_alphak} can be divided into the two cases $\alpha_k = 0$ and $\alpha_k = 1$, respectively. In the case of $\alpha_k = 0$, the RHS of \eqref{eq:proof_message_I_vkmq_alphak} is equal to 
 $a$ given in \eqref{eq:message_I_vkmq_alphak_lambda_a}. 
 In the case of $\alpha_k = 1$, the RHS of \eqref{eq:proof_message_I_vkmq_alphak} is equal to  
 \begin{subequations}
 \begin{align}
   & \int \!{\sum}_{s_{k,m,q}\in\{0,1\}} \! \mathcal{CN}\! \left( b_{k,m,q} ; \mu_{ b_{k,m,q} \to v_{k,m,q} }, \sigma^2_{ b_{k,m,q} \to v_{k,m,q} } \right) \notag\\
   & \;\;\;\; \times 
   {\rm{Ber}} \left( s_{k,m,q} ; \xi  \right) \left( (1-  s_{k,m,q}) \delta \left(   b_{k,m,q} \right)  \right. \notag\\
   & \;\;\;\;  \left.   
     + s_{k,m,q} \mathcal{CN}\left( b_{k,m,q} ; 0 , \sigma^2_{\rho,k,m,q} \right)   \right) 
   {\rm d} b_{k,m,q} \notag \\
   & = \left( 1 \!-\! \xi  \right) a  
   +  \xi \!\!\int \!\! \mathcal{CN}\! \left( b_{k,m,q} ; \mu_{ b_{k,m,q} \to v_{k,m,q} }, \sigma^2_{ b_{k,m,q} \to v_{k,m,q} } \right)
      \notag\\
   & \;\;\;\;   \times  
   \mathcal{CN} \!\left( b_{k,m,q} ; 0 , \sigma^2_{\rho,k,m,q} \right)
   {\rm d} b_{k,m,q} \label{eq:proof_message_I_vkmq_alphak_term2_1} \\ 
   & =  \left( 1 - \xi  \right)  a   +  \xi  b , \label{eq:proof_message_I_vkmq_alphak_term2}
 \end{align}
 \end{subequations}
 where \eqref{eq:proof_message_I_vkmq_alphak_term2} follows by applying the Gaussian multiplication rule with $b$ given in \eqref{eq:message_I_vkmq_alphak_lambda_b}. 
 Together with \eqref{eq:proof_message_I_vkmq_alphak}, the normalized message becomes
 \begin{subequations}
 \begin{align}
   \mathcal{M}_{ v_{k,m,q} \to \alpha_k }\! \left( \alpha_k \right)  \!& =\! (1\!-\!\alpha_k) a 
   + \alpha_k  \left(  \left( 1 \!-\! \xi  \right)  a   \!+\!  \xi  b  \right) \label{eq:proof_message_I_vkmq_alphak_lambda_1} \\
   & \propto {\rm{Ber}}\left( \alpha_k ; \lambda_{ v_{k,m,q} \to \alpha_k }  \right)   ,  \label{eq:proof_message_I_vkmq_alphak_lambda}
 \end{align}
 \end{subequations}
 where $\lambda_{ v_{k,m,q} \to \alpha_k }$ is given in \eqref{eq:message_I_vkmq_alphak_lambda}. 
 }

\section{Proof of \eqref{eq:PFIM_0}} \label{Appendix_proof_CRB}

{\blue 
The conditional PDF $p(\mathbf{y}|\bm{\psi})$ is given by
\begin{equation}\label{eq:PFIM0_py}
  p(\mathbf{y}|\bm{\psi}) 
  = \tfrac{1}{(\pi\sigma_z^2)^{LNQ}} \exp\left\{ - \tfrac{ (\mathbf{y} - \bm{\mu} )^H (\mathbf{y} - \bm{\mu} ) }{ \sigma_z^2 } \right\} ,
\end{equation}
where $\bm{\mu}\in\mathbb{C}^{LNQ}$ is obtained by stacking $\mu_{l,n,q}$'s with $\mu_{l,n,q}$ given by  
\begin{equation}\label{eq:PFIM0_py_mu_lnq}
  \mu_{l,n,q} \!= \! {\sum}_{k,m}  \alpha_{k} \bar{x}_{k,l}
  s_{k,m,q} \rho_{k,m,q}  e^{-j\pi(n-1)\sin(\theta_{m,q})} . 
\end{equation} 

Following \cite[Lemma 1]{ref:Teng-Yuan}, the $(i,i')$-th element of the FIM matrix $\mathbf{F}_{\bm{\psi},0}$ is expressed as
\begin{equation} \label{eq:PFIM0_ii}
  [\mathbf{F}_{\bm{\psi},0}]_{i,i'} 
  = \tfrac{2}{ \sigma_z^2 } {\sum}_{l,n,q} \mathbb{E}_{\bm{\psi}} \left[ \Re \left\{
  \tfrac{\partial \mu_{l,n,q}    }{\partial \psi_i} 
  \left( \tfrac{\partial \mu_{l,n,q}   }{\partial \psi_{i'}}   \right)^{*}  
  \right\}
  \right] . 
\end{equation}
We can obtain that 
\begin{equation}
  \tfrac{\partial \mu_{l,n,q}  }{\partial x_{{\rm t},m} }
  = c_{l,m,q} [\dot{\mathbf{V}}_q]_{m,n} [\mathbf{r}_{m,q}]_1, \label{eq:PFIM0_py_mu_lnq_partialx} 
\end{equation}
\begin{equation}
  \tfrac{\partial \mu_{l,n,q}  }{\partial y_{{\rm t},m} }
  = c_{l,m,q} [\dot{\mathbf{V}}_q]_{m,n} [\mathbf{r}_{m,q}]_2,  \label{eq:PFIM0_py_mu_lnq_partialy} 
\end{equation}
where 
\begin{equation} \label{eq:PFIM0_py_mu_lnq_partialV}
  [\dot{\mathbf{V}}_q]_{m,n} = -j\pi(n-1) e^{-j\pi(n-1)\sin(\theta_{m,q})},
\end{equation}
and $\mathbf{r}_{m,q}$ is given in \eqref{eq:PFIM_0_F11bar_rmq}.  
Let $\mathbf{c}_{l,q} = [c_{l,1,q},\ldots,c_{l,M,q}]^T$ and $\mathbf{R}_q = \operatorname{blkdiag}\{ \mathbf{r}_{1,q}, \ldots, \mathbf{r}_{M,q} \}$, we have 
\begin{align} \label{eq:PFIM0_py_mu_lnq_partialp}
  \tfrac{\partial \mu_{l,n,q}  }{\partial \mathbf{p}_{{\rm t}} }
  = \mathbf{R}_q ( \mathbf{c}_{l,q} \odot [\dot{\mathbf{V}}_q]_{:,n} ). 
\end{align} 

The terms $\tfrac{\partial \mu_{l,n,q}  }{\partial {\bm \rho}^{\rm R} } $ and $\tfrac{\partial \mu_{l,n,q}  }{\partial {\bm \rho}^{\rm I} }$ satisfy
\begin{equation}
  \tfrac{\partial \mu_{l,n,q}  }{\partial {\bm \rho}^{\rm R} }
  = -j \tfrac{\partial \mu_{l,n,q}  }{\partial {\bm \rho}^{\rm I} }
  = \left[ \bm{0}, (\mathbf{B}_{l,q} [\mathbf{V}_q]_{:,n} )^T, \bm{0} \right]^T ,  \label{eq:PFIM0_py_mu_lnq_partialrhoR_all}
\end{equation}
where $\mathbf{B}_{l,q} =  \operatorname{blkdiag}\{\bm{\beta}_{1,l,q}, \ldots, \bm{\beta}_{M,l,q}\}$, 
$\bm{\beta}_{m,l,q} \!=\! \bm{\alpha} \odot [\bar{\mathbf{X}}]_{l,:}^T \odot \mathbf{s}_{m,q}$, 
$\bm{\alpha} = [\alpha_1, \ldots, \alpha_K]^T$,
and $\mathbf{s}_{m,q} = [s_{1,m,q}, \ldots, s_{K,m,q}]^T$. 

Then, we can obtain that
\begin{subequations}
\begin{align} 
  \mathbf{F}_{\bm{\psi},0} 
  & = \tfrac{2}{ \sigma_z^2 } {\sum}_{l,n,q} \mathbb{E}_{\bm{\psi}} \left[ \Re \left\{
  \tfrac{\partial \mu_{l,n,q}    }{\partial \bm{\psi}} 
  \left( \tfrac{\partial \mu_{l,n,q}   }{\partial \bm{\psi}}   \right)^{H}  
  \right\}
  \right] \label{eq:F_psi0_1} \\ 
    & = \!\frac{2}{ \sigma_z^2 }  \mathbb{E}_{\bm{\psi}} \!\!\begin{bmatrix} 
        \!\Re(\mathbf{F}_{11})  & \!\!\Re(\mathbf{F}_{12}) & -\!\Im(\mathbf{F}_{12}) \\ 
        \!\Re(\mathbf{F}_{12})^T & \!\!\Re(\mathbf{F}_{22}) & -\!\Im(\mathbf{F}_{22}) \\
        -\!\Im(\mathbf{F}_{12})^T  & -\!\!\Im(\mathbf{F}_{22})^T & \!\Re(\mathbf{F}_{22}) 
      \end{bmatrix} \!, \label{eq:F_psi0_2}
\end{align}
\end{subequations}
where
\begin{equation} \label{eq:proof_F11}
  \mathbf{F}_{ 11 }
  = {\sum}_{l,q} \mathbf{R}_q ( ( \mathbf{c}_{l,q}  \mathbf{c}_{l,q}^H ) \odot
  ( \dot{\mathbf{V}}_q  \dot{\mathbf{V}}_q^H ) ) \mathbf{R}_q^H,
\end{equation}
\begin{equation} \label{eq:proof_F12}
  \mathbf{F}_{ 12 } 
  = [ \mathbf{F}_{ 12 , 1 } , \ldots ,  \mathbf{F}_{ 12 , Q } ] ,
\end{equation}
\begin{equation} \label{eq:proof_F12q}
  \mathbf{F}_{ 12 , q } = {\sum}_{l,n} 
  \mathbf{R}_q ( \mathbf{c}_{l,q} \odot [\dot{\mathbf{V}}_q]_{:,n} )  
  (\mathbf{B}_{l,q} [\mathbf{V}_q]_{:,n} )^H ,
\end{equation} 
\begin{equation} \label{eq:proof_F22}
  \mathbf{F}_{ 22 } 
  = \operatorname{blkdiag} \{ \mathbf{F}_{ 22 , 1 } , \ldots ,  \mathbf{F}_{ 22 , Q } \} ,
\end{equation} 
\begin{equation} \label{eq:proof_F22q}
  \mathbf{F}_{ 22 , q } 
  = {\sum}_l  \mathbf{B}_{l,q} \mathbf{V}_q \mathbf{V}_q^H \mathbf{B}_{l,q}^H  .
\end{equation}

Taking the expectation over $\bm{\rho}^{\rm R}$ and $\bm{\rho}^{\rm I}$, we have $\mathbb{E}_{\bm{\rho}^{\rm R},\bm{\rho}^{\rm I}} \left[\mathbf{F}_{ 12,q }\right] = {\bm 0}$, $\mathbb{E}_{\bm{\rho}^{\rm R},\bm{\rho}^{\rm I}} \left[\mathbf{F}_{ 22,q }\right] = \mathbf{F}_{ 22,q }$, and $\mathbb{E}_{\bm{\rho}^{\rm R},\bm{\rho}^{\rm I}} \left[\mathbf{F}_{ 11 }\right] = \bar{\mathbf{F}}_{ 11 }$ with the expression of $\bar{\mathbf{F}}_{ 11 }$ given in \eqref{eq:PFIM_0_F11bar}. 
}

\end{document}